\documentclass[10pt]{article} 

\usepackage[utf8]{inputenc}
\usepackage{amsmath}
\usepackage{amsfonts}
\usepackage{amssymb}
\usepackage{mathtools}
\usepackage{listings}
\usepackage{mathrsfs}
\usepackage{times}
\usepackage{float}
\usepackage{color}
\usepackage{setspace}
\usepackage{color,soul}

\usepackage{amsmath,amsfonts,amsthm,eucal}
\usepackage{enumerate}
\usepackage[letterpaper,margin=3cm]{geometry}
\usepackage{float}
\usepackage[title]{appendix}
\usepackage{color}
\usepackage{tabularx}
\usepackage{siunitx}
\usepackage{empheq}
\usepackage[tableposition=below]{caption}
\usepackage[
pdfstartview=XYZ,
bookmarks=true,
colorlinks=true,
linkcolor=blue,
urlcolor=blue,
citecolor=blue,
pdftex,
bookmarks=true,
linktocpage=true,   
hyperindex=true
]{hyperref}

\usepackage{lipsum}

\usepackage{natbib}
\usepackage[pdftex]{graphicx}
\usepackage{epstopdf}
\usepackage{booktabs} 
\usepackage{tikz,mathpazo}
\usetikzlibrary{shapes.geometric, arrows}
\usepackage{caption}

\usepackage{longtable}
\usepackage{adjustbox}
\usepackage{framed}

\usepackage[colorinlistoftodos]{todonotes}

\usepackage{listings} 
\usepackage{showexpl} 
\usepackage{color} 
\usepackage{amsthm} 
\usepackage{alltt}

\usepackage[ruled,vlined]{algorithm2e}
\usepackage{algpseudocode}
\usepackage{amssymb}
\usepackage{amsmath}
\usepackage{amsfonts}
\usepackage{booktabs}
\usepackage{enumerate}
\usepackage{fancyhdr}
\usepackage{float}
\usepackage{graphicx}
\usepackage{lastpage}
\usepackage{mathrsfs}
\usepackage{hyperref}
\hypersetup{
    colorlinks=true,
    linkcolor=blue,
    filecolor=magenta,      
    urlcolor=cyan,
}

\usepackage{appendix}
\usepackage{tikz}
\usetikzlibrary{decorations.text}
\usetikzlibrary{shapes,arrows}
\usepackage{pgfplots}
\usepackage{pgfplotstable}
\pgfplotsset{compat=newest}

\usetikzlibrary{calc}
\usetikzlibrary{spy}
\usetikzlibrary{positioning}
\usepackage{pdftexcmds}
\usetikzlibrary{external}
\usetikzlibrary{positioning}
\usetikzlibrary{arrows}
\usetikzlibrary{decorations.markings}
\usetikzlibrary{decorations.text}
\usetikzlibrary{backgrounds}
\usetikzlibrary{spy}
\usetikzlibrary{calc,patterns,decorations.pathmorphing,decorations.markings}
\usetikzlibrary{decorations.pathreplacing}
\usepgfplotslibrary{patchplots}

\usepackage{chngcntr}
\usepackage{etoolbox}
\usepackage{lipsum}

\usepackage{bm}

\makeatletter
\providecommand*{\input@path}{}
\g@addto@macro\input@path{{./}}
\makeatother

\makeatletter

\def\pgfplotstableread@openfile{%
    \def\pgfplotstable@loc@TMPa{\pgfutil@in@{ }}%
    \expandafter\pgfplotstable@loc@TMPa\expandafter{\pgfplotstableread@filename}%
    \ifpgfutil@in@
        \t@pgfplots@toka=\expandafter{\pgfplotstableread@filename}%
        \edef\pgfplotstableread@filename{\pgfplots@dquote\the\t@pgfplots@toka\pgfplots@dquote}%
    \fi
    \let\pgfplotstableread@old@crcr=\\%
    \def\\{\string\\}
    \openin\r@pgfplots@reada=\csname pgfk@/pgfplots/table file path\endcsname\pgfplotstableread@filename.tex
    \ifeof\r@pgfplots@reada
        \openin\r@pgfplots@reada=\csname pgfk@/pgfplots/table file path\endcsname\pgfplotstableread@filename\relax
    \else
        \pgfplots@warning{%
            You requested to open table '\pgfplotstableread@filename', but there is also a '\pgfplotstableread@filename.tex'. 
            TeX will automatically append the suffix '.tex', so I will now open '\pgfplotstableread@filename.tex'.
            Please make sure you don't accidentally load TeX files - this may produce unrecoverable errors.}%
        \closein\r@pgfplots@reada
        \openin\r@pgfplots@reada=\pgfplotstableread@filename\relax
    \fi
    \ifeof\r@pgfplots@reada
        \pgfplotsthrow{no such table file}{\pgfplots@loc@TMPa}{\pgfplotstableread@filename}{Could not read table file '\csname pgfk@/pgfplots/table file path\endcsname\pgfplotstableread@filename'. In case you intended to provide inline data: maybe TeX screwed up your end-of-lines? Try `row sep=crcr' and terminate your lines with `\string\\' (refer to the pgfplotstable manual for details)}\pgfeov%
        \global\let\pgfplotstable@colnames@glob=\pgfplots@loc@TMPa
        \def\pgfplotstableread@ready{0}%
    \fi
    \pgfplots@logfileopen{\pgfplotstableread@filename}%
    \let\\=\pgfplotstableread@old@crcr
}

\makeatother

\pgfplotscreateplotcyclelist{blue to red}{%
  color=blue\\%
  color=red!10!blue\\%
  color=red!20!blue\\%
  color=red!30!blue\\%
  color=red!40!blue\\%
  color=red!50!blue\\%
  color=red!60!blue\\%
  color=red!70!blue\\%
  color=red!80!blue\\%
  color=red!90!blue\\%
  color=red\\%
}
\pgfplotscreateplotcyclelist{red to blue}{%
  color=red\\%
  color=red!90!blue\\%
  color=red!80!blue\\%
  color=red!70!blue\\%
  color=red!60!blue\\%
  color=red!50!blue\\%
  color=red!40!blue\\%
  color=red!30!blue\\%
  color=red!20!blue\\%
  color=red!10!blue\\%
  color=blue\\%
}

\pgfplotscreateplotcyclelist{red to blue fast}{%
  color=red\\%
  color=red!80!blue\\%
  color=red!60!blue\\%
  color=red!40!blue\\%
  color=red!20!blue\\%
  color=blue\\%
}

\pgfplotscreateplotcyclelist{2red to blue}{%
  color=red\\%
  color=red\\%
  color=red!90!blue\\%
  color=red!90!blue\\%
  color=red!80!blue\\%
  color=red!80!blue\\%
  color=red!70!blue\\%
  color=red!70!blue\\%
  color=red!60!blue\\%
  color=red!60!blue\\%
  color=red!50!blue\\%
  color=red!50!blue\\%
  color=red!40!blue\\%
  color=red!40!blue\\%
  color=red!30!blue\\%
  color=red!30!blue\\%
  color=red!20!blue\\%
  color=red!20!blue\\%
  color=red!20!blue\\%
  color=red!10!blue\\%
  color=blue\\%
  color=blue\\%
}

\pgfplotsset{discard if/.style 2 args={x filter/.code={\ifnum\thisrow{#1}=#2\else\fi}}}

\usepackage{bm}

\newcommand{\Gammab}{{\boldsymbol \Gamma}}

\newcommand{\Sigmab}{\boldsymbol \Sigma}

\newcommand{\thetab}{\boldsymbol \theta}

\newcommand{\Fb}{{\boldsymbol F}}

\newcommand{\Nb}{{\boldsymbol N}}
\newcommand{\Pb}{{\boldsymbol P}}
\newcommand{\Qb}{{\boldsymbol Q}}

\newcommand{\Tb}{{\boldsymbol T}}
\newcommand{\Ub}{{\boldsymbol U}}
\newcommand{\Vb}{{\boldsymbol V}}
\newcommand{\Wb}{{\boldsymbol W}}

\newcommand{\bb}{{\boldsymbol b}}
\newcommand{\cb}{{\boldsymbol c}}

\newcommand{\fb}{{\boldsymbol f}}

\newcommand{\vb}{{\boldsymbol v}}

\newcommand{\yb}{{\boldsymbol y}}
\newcommand{\zb}{{\boldsymbol z}}

\newcommand{\ee}{{ \bf e}}

\newcommand{\x}{{\boldsymbol x}}
\newcommand{\y}{{\boldsymbol y}}
\newcommand{\X}{{\boldsymbol X}}

\usepackage{amsmath}

\usepackage{nicefrac}
\usepackage{multirow}
\usepackage{hhline}
\usepackage{lineno}
\usepackage{import}
\usepackage{epstopdf}
\usepackage{subcaption}
\usepackage{mathtools}
\usepackage{transparent}

\usepackage{listings}
\usepackage{cancel}
\usepackage{empheq}

\definecolor{tbf}{RGB}{255,0,0} 
\definecolor{txue}{RGB}{0,0,255}

\usepackage{authblk}

\title{Learning the nonlinear dynamics of soft mechanical metamaterials with graph networks} 

\begin{document}






\author[1]{\normalsize Tianju Xue}
\author[1]{\normalsize Sigrid Adriaenssens}
\author[2]{\normalsize Sheng Mao \footnote{\textit{maosheng@pku.edu.cn}  (corresponding author)}}

\affil[1]{\footnotesize Department of Civil and Environmental Engineering, Princeton University, Princeton, NJ 08544, USA.}
\affil[2]{\footnotesize Department of Mechanics and Engineering Science, BIC-ESAT, College of Engineering, Peking University, Beijing 100871, PRC.}

\date{}
\maketitle

\vspace{-30pt}

\begin{abstract}
The dynamics of soft mechanical metamaterials provides opportunities for many exciting engineering applications.
Previous studies often use discrete systems, composed of rigid elements and nonlinear springs, to model the nonlinear dynamic responses of the continuum metamaterials.
Yet it remains a challenge to accurately construct such systems based on the geometry of the building blocks of the metamaterial.
In this work, we propose a machine learning approach to address this challenge.
A metamaterial graph network (MGN) is used to represent the discrete system, where the nodal features contain the positions and orientations the rigid elements, and the edge update functions describe the mechanics of the nonlinear springs.
We use Gaussian process regression as the surrogate model to characterize the elastic energy of the nonlinear springs as a function of the relative positions and orientations of the connected rigid elements.
The optimal model can be obtained by ``learning'' from the data generated via finite element calculation over the corresponding building block of the continuum metamaterial.
Then, we deploy the optimal model to the network so that the dynamics of the metamaterial at the structural scale can be studied.
We verify the accuracy of our machine learning approach against several representative numerical examples.
In these examples, the proposed approach can significantly reduce the computational cost when compared to direct numerical simulation while reaching comparable accuracy.
Moreover, defects and spatial inhomogeneities can be easily incorporated into our approach, which can be useful for the rational design of soft mechanical metamaterials.  
\end{abstract}

\section{Introduction}
\label{Sec:Introduction}

Mechanical metamaterials (MMs) are commonly used in engineering for tailoring unique mechanical properties as their mechanics are primarily governed by their geometry rather than compositions.
Recent development of fabrication technology such as additive manufacture has stimulated the use of soft MMs to realize extreme mechanical properties such as negative Poisson's ratio~\citep{bertoldi2010negative, overvelde2012compaction, overvelde2014relating}, shape morphing~\citep{mirzaali2018shape}, tunable band structures~\citep{krishnan2009optical}, energy absorption~\citep{meza2014strong}.
These unique properties opened up the possibility of many exciting engineering applications, e.g., soft actuators, materials with \emph{in situ} tunable functionalities, reusable energy-absorbing materials~\citep{li2016mechanical, florijn2016programmable, bertoldi2017flexible, bertoldi2017harnessing, barchiesi2019mechanical, surjadi2019mechanical}, etc.
Furthermore, soft MMs created new ways of manipulating elastic waves with finite amplitude, that are different from the granular systems, where contact is the main mechanism~\citep{tournat2010acoustics, nesterenko2013dynamics, theocharis2013nonlinear}.

\begin{figure}[t]
    \centering 
    \scalebox{0.95}{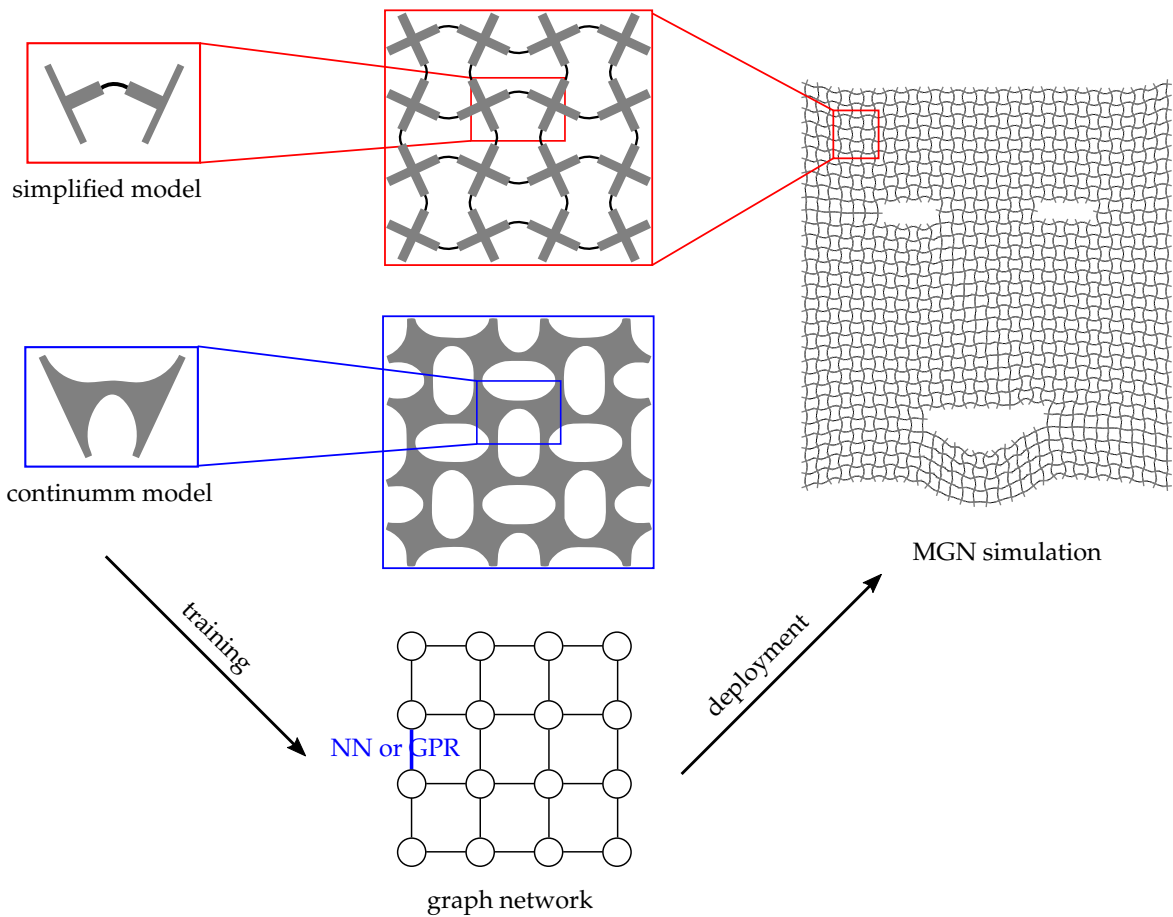}
    \caption{An overview of the proposed MGN based simulation.
    Ground truth data are generated by solving local continuum problems (blue window) so that the edge update function in the graph network can be trained with surrogate models like neura networks or Gaussian process regression.
    Red windows show the simplified system that are made of crosses connected by springs.
    The rightmost  ``smiling face'' is the results produced by a MGN based simulation, showing the deformed configuration a cross-spring system subjected to a compression in the horizontal direction.}
    \label{Fig:abstract}
\end{figure}

Many of the unique properties of soft MMs mentioned above are determined by their dynamic behavior, which is often nonlinear.
Initial studies have focused on systems where nonlinearity occurs at the quasi-static stage with a linearized dynamics~\citep{bertoldi2008wave, shan2014harnessing, goldsberry2018negative}.
The nonlinear dynamics of soft MMs as continuum remains a challenge, as the 
nonlinearity is oftentimes the combination of the geometric nonlinearity due to large deformation, material nonlinearity due to the constitutive relations of the solid constituents, and those caused by possible mechanical instabilities.
For such reason, many studies have focused on the use of simplified discrete systems and one widely adopted such system is shown in Fig.~\ref{Fig:abstract}, where the continuum MM is replaced by a system of rigid elements connected by nonlinear springs~\citep{deng2021nonlinear}.
The rigid elements can translate and rotate, and the nonlinear springs can support both longitudinal forces as well as torques.
Such systems can greatly reduce the number of degrees of freedom (DOFs) needed for numerical calculation while still provide essential insights into important dynamic responses of soft MMs such as cnoidal waves~\citep{deng2021dynamics}, vector solitons~\citep{deng2017elastic,deng2019focusing}, rarefaction solitons~\citep{yasuda2019origami} and topological solitons~\citep{hussein2014dynamics}, etc.
Although certain mechanical model of the nonlinear springs in the simplified systems works well with soft MMs of some unit-cell geometry, it is still largely an unanswered question how such model can be constructed accurately for MMs with complex unit-cell geometry, which is vital for the rational design of soft MMs for desired dynamic responses.
In this work, we make one step towards establishing such connection with the help of machine learning methods, and specifically, the \textit{graph networks}.

Generally speaking, graph networks are a class of machine learning methods that operate on graph-like data structures.
As noted in~\citep{mitchell1980need,battaglia2018relational}, graph networks ``carry strong relational inductive biases, in the form of specific architectural assumptions, which guide these approaches towards learning about entities and relations''.
Although the term ``graph neural networks'' may be more well-known, the functions that graph networks define do not need to be neural networks.
In fact they can be any functions in principle.
Graph networks have found its applications in many scientific and engineering problems, such as visual scene understanding~\citep{santoro2017simple}, reasoning about knowledge graphs~\citep{bordes2013translating}, predicting chemical properties of molecules~\citep{duvenaud2015convolutional}, predicting traffic on roads~\citep{li2017diffusion}, etc.
Lately, we also observe a surging trend of applying graph networks in computational mechanics.
Several architectures and methods regarding graph networks were proposed for leaning and inference of dynamics of physical systems~\citep{battaglia2016interaction,greydanus2019hamiltonian,sanchez2018graph,chang2016compositional}.
Pfaff et al.~\citep{pfaff2020learning} made use of the graph nature of computational meshes used in many numerical algorithms like the finite element method (FEM)~\citep{hughes2012finite} and used graph neural networks to accelerate mesh-based simulations.
By learning particle relations from particle-based methods such as ``smoothed particle hydrodynamics'' (SPH)~\citep{monaghan1992smoothed} and ``material point method'' (MPM)~\citep{sulsky1995application}, Sanchez-Gonzalez et al.~\citep{sanchez2020learning} designed simulators based on graph neural networks that have comparable numerical performance.
Graph data structures were employed to represent anisotropic constitutive relations of hyperelasticity~\citep{vlassis2020geometric} and elastoplasticity~\citep{heider2020so}.
Mozaffar et al.~\citep{mozaffar2021geometry} applied a graph-based representation to capture spatiotemporal dependencies of thermal responses in additive manufacturing processes.

In the present work, we propose a graph network based framework, which we refer to as the metamaterial graph network (MGN), to model the nonlinear dynamics of soft MMs.
As shown in Fig.~\ref{Fig:abstract}, MGN replaces the continuum MM with a system of rigid crosses connected by nonlinear springs, which can be regarded as a graph, whose nodes represent the rigid crosses and connectivities between the nodes represent the nonlinear springs.
The kinematics of the rigid crosses can be readily described by Newton's law, and the key of this framework is to construct a model of the nonlinear springs to best represent the mechanics of the given MM.
To achieve this goal, we first generate training data by conducting finite element simulation of one building block of the MM.
In the simulation, we alter the configuration of the building block, which reflects the changes in the relative positions and orientations of the connecting crosses in the discrete system.
Then, we use Gaussian process regression (GPR) as the surrogate model of the nonlinear spring to learn the elastic energy of the building block, which is a function of the relative positions and orientation of the connected crosses.
We also use neural networks (NNs) as our surrogate model for comparison and in the problems that we are concerned with, GPR exhibits apparent advantages over NNs.
By supplying the MGN with the trained GPR model, the dynamics MMs can be obtained with a much lower computational cost compared to direct numerical simulation (DNS).

In this work, we carry out the finite element simulation using the open-source package \texttt{FEniCS}~\citep{logg2012automated}, while training of the surrogate models as well as numerical simulation of the dynamics of the cross-spring system are both implemented in \texttt{JAX}~\citep{jax2018github}, a Python library designed for high-performance numerical computing with GPU support.
While \texttt{JAX} is mostly used within the machine learning community, its potential in scientific computing has been revealed by several recent works, e.g., molecular dynamics simulation~\citep{schoenholz2020jax} or computational fluid dynamics~\citep{kochkov2021machine}.
Like those works, we make full use of the powerful features provided by \texttt{JAX}, e.g., automatic differentiation (\texttt{grad}), vectorization (\texttt{vmap}), and just-in-time compilation (\texttt{jit}) to further enhance the performance of our MGN based simulation.
Our code is available at \href{https://github.com/tianjuxue/gnmm}{https://github.com/tianjuxue/gnmm}.

The paper is organized as follows.
Section~\ref{Sec:cmm} provides a brief overview on the continuum theory of the dynamics of soft MMs, and Section~\ref{Sec:cs} describes a cross-spring system as a simplified model to approximate the continuum metamaterials.
In Section~\ref{Sec:gn}, we briefly introduce the fundamentals of graph networks, including a brief overview of the surrogate models used in the graph networks.
Section~\ref{Sec:construction} discusses the training, validation and testing of the surrogate models.
In Section~\ref{Sec:numerical}, we present several numerical examples to show that our graph network based simulation can effectively capture the nonlinear dynamics of different soft MMs.
We then conclude in Section~\ref{Sec:conc}.

\section{Elastodynamics of continuum metamaterials}
\label{Sec:cmm}


\subsection{Elastodynamics under finite deformation}

Consider an elastic body that occupies a region ~${\mathbb{B} \subset \mathbb{R}^3}$, at its stress-free reference state.
We denote by $\partial \mathbb{B}$ the boundary of the body and $\Nb$ the outward normal of $\partial \mathbb{B}$.
The boundary $\partial \mathbb{B}$ is assumed to be composed of two parts $\partial \mathbb{B}_D$ and $\partial \mathbb{B}_N$ such that~${\partial\mathbb{B}_N \cup \partial\mathbb{B}_D = \partial \mathbb{B}}$ and~${\partial\mathbb{B}_N \cap \partial\mathbb{B}_D = \emptyset}$.
We consider the dynamics of the body over a time period $[0, T]$.
For any time $t\in[0,T]$, the body deforms and occupies a spatial region~${\mathbb{B}_t \subset \mathbb{R}^3}$.
The deformation map~${\boldsymbol{\varphi}_t: \mathbb{B} \times [0,T] \rightarrow \mathbb{B}_t}$ sends a material point~${\X \in \mathbb{B}}$ to its spatial counterpart~${\x \in \mathbb{B}_t}$, i.e.,~${\x = \boldsymbol{\varphi}(\X, t)}$. 
The corresponding displacement field is defined as~${\Ub(\X, t) = \x - \X}$, and the velocity $\dot{\Ub}=\frac{\partial \Ub}{\partial t}$.
The total action functional of the body is
\begin{align} \label{Eq:action}
    S(\Ub) = \int_{0}^{T} L(\Ub, \dot{\Ub}) \textrm{d}t,
\end{align}
under some initial and boundary conditions. 
Here, we prescribe initial conditions for displacement $\Ub(\X, 0)=\Ub_0: \mathbb{B} \rightarrow \mathbb{R}^3$, and velocity $\dot{\Ub}(\X, 0) = \Vb_0:\mathbb{B} \rightarrow \mathbb{R}^3$.
We also prescribe displacement $\Ub_D: \partial \mathbb{B}_D \times [0, T] \rightarrow \mathbb{R}^3$, and traction $\Tb_N: \partial \mathbb{B}_N \times [0, T] \rightarrow \mathbb{R}^3$ boundary conditions. 
Both of them can be time dependent.
For a hyperelastic material (in the absence of body force), the Lagrangian can be written as~\citep{marsden1994mathematical} 

\begin{align} \label{Eq:lagrangian}
   L(\Ub, \dot{\Ub}) = \int_{\mathbb{B}} \frac{1}{2}\rho_R |\dot{\Ub}|^2  \textrm{ d}\X - \int_{\mathbb{B}} W(\Fb) \textrm{ d}\X +  \int_{\partial\mathbb{B}_{N}} \Tb_N \cdot \Ub \textrm{ d} \Gammab,
\end{align}
where $\rho_R$ is the mass density in the reference configuration, ${\Fb = \frac{\partial \x}{\partial \X} }$ is the deformation gradient, and $W$ is the strain energy density function (per volume).
The Euler-Lagrange equations corresponding to minimizing Eq.~(\ref{Eq:action}) are
\begin{align} \label{Eq:BVP-strong}
   \rho_R\frac{\partial^2 \Ub}{\partial t^2} = \textrm{Div }  \Pb & \quad \textrm{in}  \, \, \mathbb{B} \times [0, T], \nonumber \\
    \Ub(\X, 0) = \Ub_0 & \quad\textrm{in} \, \, \mathbb{B},  \nonumber \\
    \dot{\Ub}(\X, 0) = \Vb_0 & \quad\textrm{in} \, \, \mathbb{B},  \nonumber \\
    \Ub = \Ub_D &  \quad\textrm{on}  \, \, \partial\mathbb{B}_D \times [0, T],  \nonumber \\
    \Pb \cdot \Nb = \Tb_N  & \quad \textrm{on} \, \, \partial\mathbb{B}_N \times [0, T],
\end{align}
where the divergence operator is defined in the reference configuration $\mathbb{B}$, and the first Piola-Kirchhoff stress $\Pb$ is given by~${\Pb = \frac{\partial W}{\partial \Fb}}$.
Throughout this work, we adopt the following strain energy density function $W$:
\begin{align} \label{Eq:neo-hookean}
    W (\Fb) = \frac{G}{2}(J^{-2/3} I_1 - 3) + \frac{\kappa}{2}(J - 1)^2,
\end{align}
where~${J = \textrm{det}(\Fb)}$,~${I_1 = \textrm{tr}(\boldsymbol{C})}$;~${G = \frac{E}{2(1+\nu)}}$ and~${\kappa = \frac{E}{3(1-2\nu)}}$ denote the initial shear and bulk moduli, respectively, with~${E}$ being the Young's modulus and and~$\nu$ the Poisson's ratio of the material.
The above~${W}$ is commonly used to model isotropic elastomers that are almost incompressible~\citep{ogden1997non, pence2015compressible}.
Moreover, plane strain is assumed here, i.e.,~${U_{i} = U_{i}(X_1, X_2, t), \ i = 1, 2}$ and~${U_{3} = 0}$. 

\begin{figure}[!ht]
    \centering 
    \scalebox{0.8}{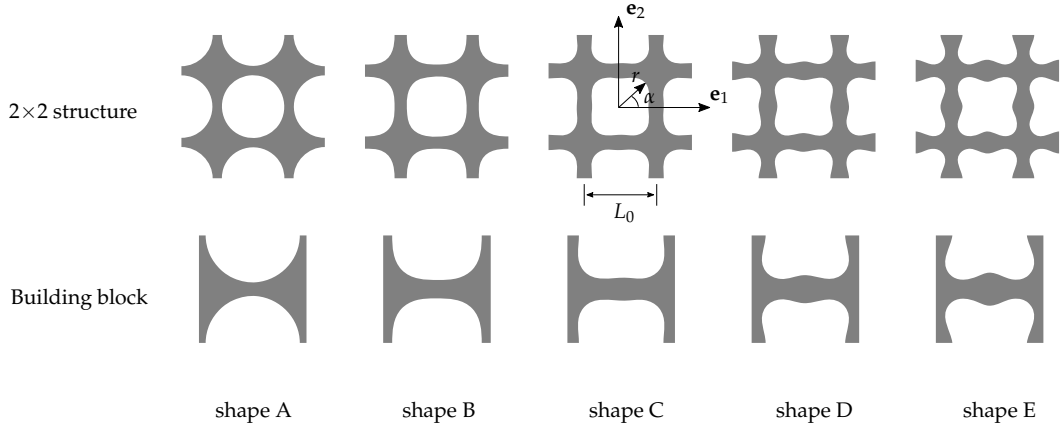}
    \caption{The five representative pore shapes. In the first row, five 2$\times$2 CMMs are shown, each with a different pore shape. The second row shows the corresponding building blocks.}
    \label{Fig:pore}
\end{figure} 

\subsection{Problem geometry}

In this work, we constrain ourselves to the study of 2D soft cellular mechanical metamaterials (CMMs) made of square unit-cells with a pore in the center.
Inspired by previous works~\citep{overvelde2012compaction, overvelde2014relating}, we consider the pore shapes with four-fold symmetry whose contour can be described by the following equation:
\begin{align} \label{Eq:pore}
	r(\alpha) &= r_0\left[1 + \xi \textrm{cos}(4\alpha)\right],
\end{align}
where, as is shown in Fig.~\ref{Fig:pore}, $r$ and $\alpha$ are polar radius and polar angle respectively; $\xi$ is a parameter that controls the shape, and $r_0$ is related to the initial porosity $\phi_0$ via:
\begin{align} \label{Eq:r0}
	r_0 = \dfrac{L_0\sqrt{2\phi_0}}{\sqrt{\pi (2 + \xi^2)}},
\end{align}
with $L_0$ the length of a single unit-cell.
A family of various pore shapes are illustrated in Fig.~\ref{Fig:pore} with $\xi=0, -0.05, -0.1, -0.15, -0.2$.
We label these five representative shapes with ``shape A'', ``shape B'', ``shape C'', ``shape D'', and ``shape E''.
In Fig.~\ref{Fig:pore}, the first row shows the 2$\times$2 structures of these five pore shapes.

In principle, the dynamic response of our soft CMMs can be obtained by DNS, i.e. numerically solving (\ref{Eq:BVP-strong}) over the whole structure.
In this work, we use finite element method (FEM) to carry out the DNS by solving the weak form of (\ref{Eq:BVP-strong}).
The classic Rothe's method~\citep{rothe1930zweidimensionale,rektorys1971application,nevcas1974application} is used for time integration.
See~\ref{App:dns} for details.


\section{Simplified cross-spring system}
\label{Sec:cs}

\begin{figure}[!ht]
    \centering 
    \scalebox{0.9}{
\begingroup%
  \makeatletter%
  \providecommand\color[2][]{%
    \errmessage{(Inkscape) Color is used for the text in Inkscape, but the package 'color.sty' is not loaded}%
    \renewcommand\color[2][]{}%
  }%
  \providecommand\transparent[1]{%
    \errmessage{(Inkscape) Transparency is used (non-zero) for the text in Inkscape, but the package 'transparent.sty' is not loaded}%
    \renewcommand\transparent[1]{}%
  }%
  \providecommand\rotatebox[2]{#2}%
  \newcommand*\fsize{\dimexpr\f@size pt\relax}%
  \newcommand*\lineheight[1]{\fontsize{\fsize}{#1\fsize}\selectfont}%
  \ifx\svgwidth\undefined%
    \setlength{\unitlength}{449.42065492bp}%
    \ifx\svgscale\undefined%
      \relax%
    \else%
      \setlength{\unitlength}{\unitlength * \real{\svgscale}}%
    \fi%
  \else%
    \setlength{\unitlength}{\svgwidth}%
  \fi%
  \global\let\svgwidth\undefined%
  \global\let\svgscale\undefined%
  \makeatother%
  \begin{picture}(1,0.608247)%
    \lineheight{1}%
    \setlength\tabcolsep{0pt}%
    \put(0,0){\includegraphics[width=\unitlength,page=1]{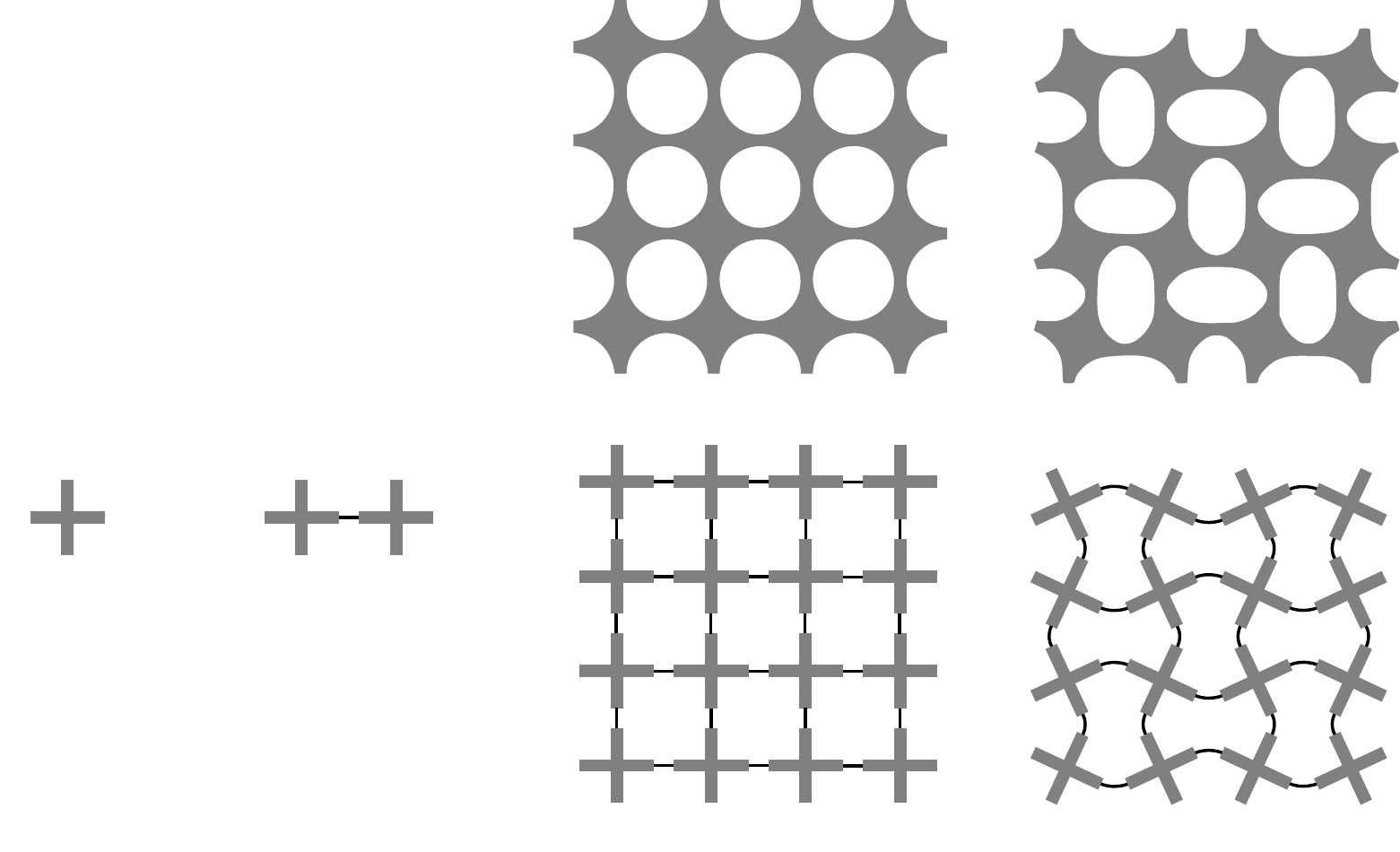}}%
    \put(0.53599854,0.311807){\color[rgb]{0,0,0}\makebox(0,0)[lt]{\lineheight{1.25}\smash{\begin{tabular}[t]{l}(a)\end{tabular}}}}%
    \put(0.85899763,0.30666795){\color[rgb]{0,0,0}\makebox(0,0)[lt]{\lineheight{1.25}\smash{\begin{tabular}[t]{l}(b)\end{tabular}}}}%
    \put(0.53438182,0.00346311){\color[rgb]{0,0,0}\makebox(0,0)[lt]{\lineheight{1.25}\smash{\begin{tabular}[t]{l}(c)\end{tabular}}}}%
    \put(0.85355889,0.00346311){\color[rgb]{0,0,0}\makebox(0,0)[lt]{\lineheight{1.25}\smash{\begin{tabular}[t]{l}(d)\end{tabular}}}}%
    \put(0,0){\includegraphics[width=\unitlength,page=2]{overview.pdf}}%
    \put(-0.00021466,0.18348811){\color[rgb]{0,0,0}\makebox(0,0)[lt]{\lineheight{1.25}\smash{\begin{tabular}[t]{l}A cross\end{tabular}}}}%
    \put(0.03488566,0.31121833){\color[rgb]{0,0,0}\makebox(0,0)[lt]{\lineheight{1.25}\smash{\begin{tabular}[t]{l}Continuum counterparts\end{tabular}}}}%
    \put(0.11332548,0.18293491){\color[rgb]{0,0,0}\makebox(0,0)[lt]{\lineheight{1.25}\smash{\begin{tabular}[t]{l}A spring with two corsses\end{tabular}}}}%
  \end{picture}%
\endgroup%
}
    \caption{The continuum porous CMMs and the simplified cross-spring system. (a) A stress-free CMM in the reference configuration. (b) A deformed CMM in the spatial configuration. (c) The cross-spring system in the reference configuration. (d) The deformed cross-spring system in the spatial configuration.}
    \label{Fig:overview}
\end{figure}

DNS of soft CMMs using FEM, although possible in principle, can easily become computationally intractable for large-size structures due to the computational cost.
To overcome this difficulty, many previous works adopt discrete systems to approximate the continuum CMM~\citep{nadkarni2014dynamics,kochmann2017exploiting,deng2017elastic,deng2019propagation,deng2021dynamics}, that are typically composed of stiffer elements connected by flexible hinges.
Such systems can greatly reduce the number of DOFs involved in the numerical calculation while still capturing the essential dynamics of the CMM.
For the type of CMMs considered in this work, since the internal rotation plays an important role, the stiffer elements are modeled as rigid crosses, that can undergo translation as well as rotation; while the flexible hinges are modeled as nonlinear springs, as shown in Fig.~\ref{Fig:overview}(a) and (c). 
These springs can provide both longitudinal forces as well as torques. 

\begin{figure}[t]
    \centering 
    \scalebox{0.9}{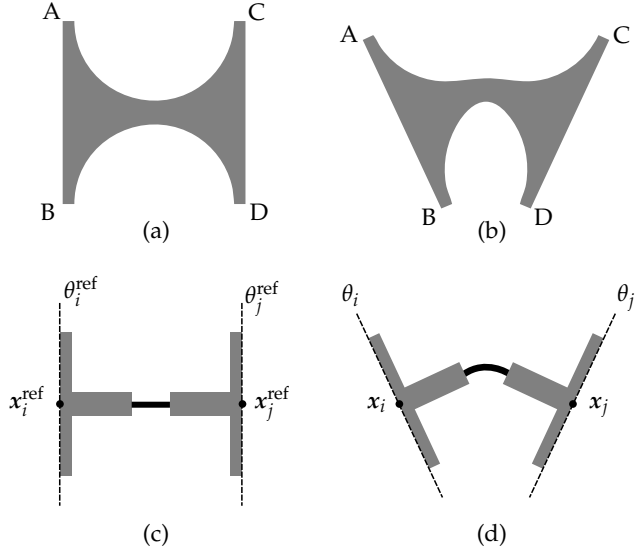}
    \caption{A building block for the continuum CMM in its reference configuration (a) and spatial configuration (b).
    The corresponding building block for the cross-spring system that consists of a spring connecting two half crosses in the reference configuration (c) and spatial configuration (d).
    In this specific example, we have $\theta_i^{\textrm{ref}}=\theta_j^{\textrm{ref}}=\frac{\pi}{2}$, $\theta_i=\frac{2\pi}{3}$, and $\theta_j=\frac{\pi}{3}$.
    }
    \label{Fig:beam}
\end{figure}

To better understand how the simplified system works, let us take a look at the building block of a continuum CMM shown in Fig.~\ref{Fig:beam}(a), which is in its reference state.
We replace such building block by two half crosses connected by a nonlinear spring, which is shown in Fig.~\ref{Fig:beam}(c).
We denote the reference positions of the two half crosses by $\x_i^{\textrm{ref}}$ and $\x_j^{\textrm{ref}}$ respectively and they correspond to the respect mid-points of sides AB and CD.
The reference orientations of the two half crosses by $\theta_i^{\textrm{ref}}$ and $\theta_j^{\textrm{ref}}$ respectively and they correspond to the respective orientations of AB and CD (with respect to a certain fixed axis).
In this state, the elastic energy stored in the spring is 0 as the CMM is stress-free.
Suppose the building block is deformed into the shape shown in Fig.~\ref{Fig:beam}(b) in its spatial configuration, the positions and the orientation of the two sides are altered.
For the simplified system, the positions and orientation of the left and right half crosses change accordingly to $(\x_i, \theta_i)$ and $(\x_j, \theta_j)$ respectively, as shown in Fig.~\ref{Fig:beam}(d).
In the meantime, the elastic energy of the building block due to deformation is completely prescribed to the spring, which is a function of $(\x_i, \theta_i)$, $(\x_j, \theta_j)$, $(\x_i^{\textrm{ref}}, \theta_i^{\textrm{ref}})$ and $(\x_j^{\textrm{ref}}, \theta_j^{\textrm{ref}})$.

Now consider a system composed of $N_c$ such crosses that are connected by $N_e$ nonlinear springs and we study its dynamics over a time period $[0, T]$.
We denote the reference position and orientation of the $i$th cross at the stress-free state by $\x_i^{\textrm{ref}}$ and $\theta_i^{\textrm{ref}}$ respectively.
In this state, the total elastic energy stored inside the springs needs to be 0.
At time $t\in[0, T]$, the spatial position and orientation of the $i$th cross are denoted by  $\x_i$ and $\theta_i$ respectively.
We further define the reference generalized coordinates as: $\Xi^{\textrm{ref}}=\lbrace\x_i^{\textrm{ref}}\rbrace_{i=1:N_c}$ and $\Theta^{\textrm{ref}}=\lbrace\theta_i^{\textrm{ref}}\rbrace_{i=1:N_c}$ and the spatial generalized coordinates as: $\Xi=\lbrace\x_i\rbrace_{i=1:N_c}$ and $\Theta=\lbrace\theta_i\rbrace_{i=1:N_c}$, with their corresponding velocities:  $\dot{\Xi}=\lbrace\dot{\x_i}\rbrace_{i=1:N_c}$ and $\dot{\Theta}=\lbrace\dot{\theta}_i\rbrace_{i=1:N_c}$.
Then the total action function of the system within time period $[0, T]$ is
\begin{align} \label{Eq:action_cs}
    S(\Xi, \Theta) = \int_{0}^{T}  L(\Xi,\dot{\Xi},\Theta,\dot{\Theta}) \textrm{d}t,
\end{align}
under some initial and boundary conditions for the generalized coordinates and velocities.
The Lagrangian of the system can be described as
\begin{align} \label{Eq:lagrangian_cs}
   L(\Xi,\dot{\Xi},\Theta,\dot{\Theta}) = \sum_{i=1}^{N_c} \frac{1}{2} m_i |\dot{\x_i}|^2 +  \sum_{i=1}^{N_c} \frac{1}{2} I_i |\dot{\theta_i}|^2  - \Psi(\Xi, \Theta, \Xi^{\textrm{ref}}, \Theta^{\textrm{ref}}) + \sum_{i=1}^{N_c} \fb_i\cdot \x_i + \sum_{i=1}^{N_c} \tau_i  \theta_i,
\end{align}
where $m_i$ and $I_i$ are the mass and moment of inertia of the $i$th cross respectively.
$\Psi$ is the total elastic energy due to the deformation of the springs; 
$\fb_i$ and $\tau_i$ are prescribed external forces and torques exerted on the $i$th cross, which are related to the boundary conditions. 
The Euler-Lagrange equations corresponding to minimizing Eq.~(\ref{Eq:action_cs}) are
\begin{align} \label{Eq:strong_cs}
    m_i\frac{\textrm{d}^2 \x_i}{\textrm{d}t^2} &= -\frac{\partial \Psi}{\partial \x_i} + \fb_i, \nonumber \\
    I_i\frac{\textrm{d}^2 \theta_i}{\textrm{d}t^2} &= -\frac{\partial \Psi}{\partial \theta_i} + \tau_i,
\end{align}
where $i=1,2,...,N_c$.
Eq.~(\ref{Eq:strong_cs}) is a system of second-order ordinary differential equations (ODEs).
Given some proper initial and boundary conditions, the dynamics of the cross-spring system is primarily determined by $\Psi$.
For some given $\Psi$, Eq.~(\ref{Eq:strong_cs}) can be numerically integrated to solve for the dynamics of the system (see \ref{App:cs} for details).
However, in general the form of $\Psi$ is not known \textit{a priori}.
While certain forms of $\Psi$ work well with CMMs with some pore geometry, it remains largely unclear how it can be constructed for CMMs with complex pore shapes, such as those shown in Fig.~\ref{Fig:pore}.
In the following section, we will show how such form can be constructed accurately with the help of graph networks.

\section{Metamaterial graph network}
\label{Sec:gn}

We introduce our metamaterial graph network (MGN) framework in this section, which is used to determine the functional form of $\Psi$.
We start by introducing the fundamentals of graph networks.
By representing the simplified cross-spring system with a graph network, $\Psi$ can be easily determined from the edge update and aggregation functions, which needs to be trained. 
Then we provide a brief overview of two classes of surrogate models for edge update functions, namely, Gaussian process regression (GPR) and neural networks (NNs).

\subsection{Fundamentals of graph network}

\begin{figure}[t]
    \centering 
    \scalebox{0.9}{
\begingroup%
  \makeatletter%
  \providecommand\color[2][]{%
    \errmessage{(Inkscape) Color is used for the text in Inkscape, but the package 'color.sty' is not loaded}%
    \renewcommand\color[2][]{}%
  }%
  \providecommand\transparent[1]{%
    \errmessage{(Inkscape) Transparency is used (non-zero) for the text in Inkscape, but the package 'transparent.sty' is not loaded}%
    \renewcommand\transparent[1]{}%
  }%
  \providecommand\rotatebox[2]{#2}%
  \newcommand*\fsize{\dimexpr\f@size pt\relax}%
  \newcommand*\lineheight[1]{\fontsize{\fsize}{#1\fsize}\selectfont}%
  \ifx\svgwidth\undefined%
    \setlength{\unitlength}{190.48661216bp}%
    \ifx\svgscale\undefined%
      \relax%
    \else%
      \setlength{\unitlength}{\unitlength * \real{\svgscale}}%
    \fi%
  \else%
    \setlength{\unitlength}{\svgwidth}%
  \fi%
  \global\let\svgwidth\undefined%
  \global\let\svgscale\undefined%
  \makeatother%
  \begin{picture}(1,0.75593517)%
    \lineheight{1}%
    \setlength\tabcolsep{0pt}%
    \put(0,0){\includegraphics[width=\unitlength,page=1]{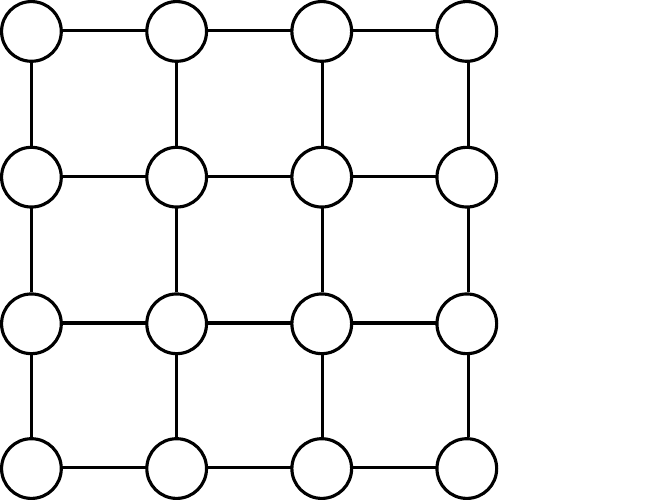}}%
    \put(0.46659865,0.47569807){\color[rgb]{0,0,0}\makebox(0,0)[lt]{\lineheight{1.25}\smash{\begin{tabular}[t]{l}$\cb_i$\end{tabular}}}}%
    \put(0.57240742,0.51876386){\color[rgb]{0,0,0}\makebox(0,0)[lt]{\lineheight{1.25}\smash{\begin{tabular}[t]{l}$e_k$\end{tabular}}}}%
    \put(0,0){\includegraphics[width=\unitlength,page=2]{graph.pdf}}%
    \put(0.91823166,0.36640608){\color[rgb]{0,0,0}\makebox(0,0)[lt]{\lineheight{1.25}\smash{\begin{tabular}[t]{l}$g$\end{tabular}}}}%
  \end{picture}%
\endgroup%
}
    \caption{Graph representation of a 4$\times$4 cross-spring system. Here, $g$ is the global feature, $\cb_i$ is the node feature associated with node $i$, and $e_k$ is the edge feature associated with (undirected) edge $k$.}
    \label{Fig:graph}
\end{figure}

The cross-spring system introduced in the previous section can be represented by a graph if each cross is regarded as a node and each spring as an edge in the graph.
For example, the 4$\times$4 cross-spring system in Fig.~\ref{Fig:overview} can be represented by an undirected graph in Fig.~\ref{Fig:graph}.
Consider a cross-spring system with $N_c$ crosses that are connected by $N_e$ springs.
Following the treatment of~\citep{battaglia2018relational}, we use a graph $\mathcal{G}$ to represent such system, which can be formally defined by a 3-tuple $\mathcal{G}=(g, V, E)$.
Here, $g$ is a global feature and in our MGN, it represents the total elastic energy of the cross-spring system $\Psi$.
The set $V=\lbrace\cb_i\rbrace_{i=1:N_c}$ is the collection of all nodal features, and in our MGN, the feature $\cb_i$ includes the reference as well as the spatial positions and orientations of the $i$th cross, i.e., $\cb_i=(\x_i^{\textrm{ref}}, \theta_i^{\textrm{ref}}, \x_i, \theta_i)$.
The set $E=\lbrace(e_k, a_k, b_k)\rbrace_{k=1:N_e}$ is the collection of edge features, namely, $e_k$ describes the feature of the edge connecting nodes $a_k$
and $b_k$.
In our MGN, we define $e_k$ as the elastic energy stored in the $k$th spring, which connects the $a_k$th and $b_k$th crosses.  
In addition to the graph structure $\mathcal{G}$, two major steps are needed for our MGN, namely \textit{update} and \textit{aggregation}:
\begin{subequations}
\begin{align}
    \text{update: } e_k &= \psi(\cb_{a_k}, \cb_{b_k}), \quad \forall k\in\lbrace 1,2,...,N_e\rbrace, \\
    \text{aggregation: }g &= \zeta(E).
\end{align}
\end{subequations}
Here, $\psi$ is often called the edge update function, and in our MGN it represents functional form of the elastic energy of the $k$th spring, which is a function of the features of node $a_k$ and $b_k$ (their generalized coordinates).
In our MGN, the aggregation function $\zeta$ is fixed such that $\zeta(E):=\sum_{k=1}^{N_e} e_k$, in order words, the total elastic energy is a summation of the elastic energy stored in each individual spring.

If $\psi$ is a given function, then computing $g$ (or $\Psi$) is straightforward.
As described in the previous section, such functional form is often not known \textit{a priori}.
In our MGN framework, we use a data-driven approach to train an approximate surrogate for $\psi$, so that $\psi$ can be ``learned'' from certain data sets.
The detailed procedures are shown in Alg.~\ref{Alg:graph}.

\begin{algorithm}[H]
\caption{Construction of $\Psi$ given ($\Xi, \Theta$) using a graph network.\label{Alg:graph}}
\textbf{input:} ($\Xi, \Theta$) \\
\For{$i\in\lbrace 1,2,...,N_c\rbrace$}{
    $\cb_i \leftarrow (\Xi_i^{\textrm{ref}}[i], \Theta_i^{\textrm{ref}}[i], \Xi[i], \Theta[i])$ \tcp{Update node features} 
}
\For{$k\in\lbrace 1,2,...,N_e\rbrace$}{
    $e_k \leftarrow \psi(\cb_{a_k}, \cb_{b_k})$ \tcp{Update edge features}
}
\textbf{let} $E \leftarrow \lbrace(e_k, a_k, b_k)\rbrace_{k=1:N_e}$ \\
\textbf{let} $g \leftarrow \zeta(E)$ \tcp{Update global feature}
\textbf{let} $\Psi \leftarrow g$ \\
\textbf{output:} $\Psi$
\end{algorithm}

\subsection{Surrogate models: neural network and Gaussian process regression}

\begin{figure}[H]
    \centering 
    \scalebox{0.8}{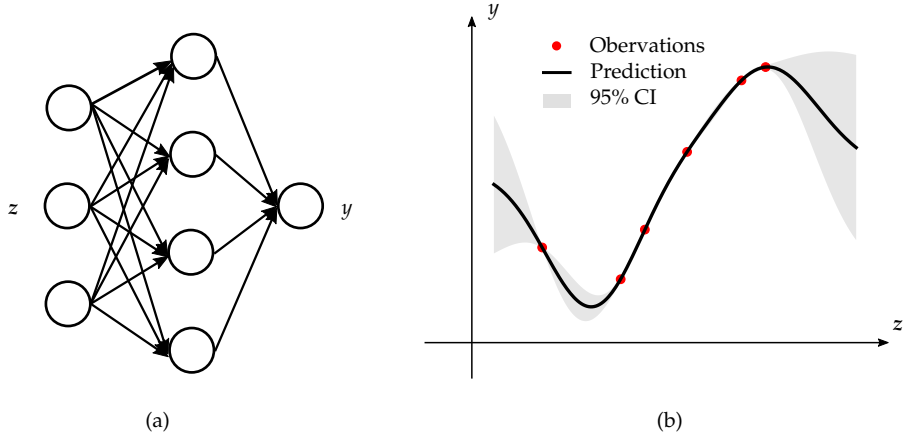}
    \caption{Schematics of the two surrogate models used in this work: (a) A multi-layer perceptron with only one hidden layer that maps the input vector $\zb$ to the output scalar $y$; (b) Gaussian process regression that predicts $y$ conditioned on observed data.}
    \label{Fig:regression}
\end{figure}

Under supervised learning, we train a surrogate model that best approximates the true ${\psi}$ over some data set.
For the type of CMMs studied in this work, ${\psi}$ can be highly nonlinear.
Therefore, simple linear regression or its direct variants like ridge regression~\citep{hoerl1970ridge} or LASSO~\citep{tibshirani1996regression} may not be ideal candidates for our surrogate models.
To deal with strongly nonlinearity, one can resort to neural networks~\citep{bishop2006pattern}.
In this work, we adopt one widely used form of neural networks, the multi-layer perceptron (MLP), which is composed of fully-connected layers~\citep{hastie2009elements}. 
The input vector $\zb$ is processed through several hidden layers and finally to an output scalar $y$.
The transformation between the~$i$th layer ($m$ neurons) and the $i+1$ layer ($n$ neurons) can be defined as:
\begin{align}
    \yb_{i} = h_{i}(\zb_{i}):= \sigma(\Wb_{i} \cdot \zb_{i} + \bb_{i}),
\end{align}
where~${\Wb_i \in \mathbb{R}^{n\times m}}$,~${\bb_i \in \mathbb{R}^{m}}$ are the weight matrix and bias vector of the~$i$th layer, and~$\sigma$ is an element-wise nonlinear function, such as a logistic function or $\tanh$. 
We can ``stack'' these transformations as~${y = h_k(\zb_k) = h_{k} \circ h_{k-1} (\zb_{k-1}) = ... =  h_k \circ ... \circ h_1 (\zb)}$ for a MLP with~$k-1$ hidden layers.
We hereby denote this composed mapping as
\begin{align}
    y = h_{\thetab} (\zb),
\end{align}
where the learnable parameter~${\thetab := (\Wb_1, \Wb_2, ..., \Wb_k; \bb_1, \bb_2, ..., \bb_k)}$.
A typical MLP with only one hidden layer is shown in Fig.~\ref{Fig:regression}(a).

MLP is a \textit{parametric} model in that it has finite number of parameters and once those parameters have been determined, training data is no longer needed when making predictions.
Alternatively, Gaussian process regression, a \textit{nonparametric} model, can also be used for the regression of highly nonlinear functions~\citep{rasmussen2003gaussian}.
In Gaussian process regression (GPR), we assume a Gaussian process prior 
\begin{align}
    f(\zb) \sim \mathcal{GP}(m(\zb), k(\zb, \zb')),
\end{align}
where the mean function $m(\zb)$ and the covariance function $k(\zb, \zb')$ fully determines the function $f(\zb)$.
In this work, the mean function $m(\zb)$ is set to be zero, and the covariance function $k(\zb, \zb')$ is set to be the commonly used squared exponential kernel
\begin{align}
    k(\zb, \zb')=\sigma^2\textrm{exp}\Big(-\frac{|\zb-\zb'|^2}{2l^2}\Big),
\end{align}
where $\sigma$ is the output variance parameter and $l$ is the length scale parameter. 
We further assume the observation
$y$ is subject to an independent additive Gaussian noise such that
\begin{align}
    y=f(\zb) + \epsilon, \quad \epsilon \sim \mathcal{N}(0, \sigma_n^2),
\end{align}
where $\mathcal{N}(0, \sigma_n^2)$ denotes the normal distribution with mean $0$ and variance $\sigma_n^2$.

Let the training data be $Z=\lbrace \zb^{(i)} \rbrace$ and $\yb=\lbrace y^{(i)} \rbrace$.
At inference phase, we aim at obtaining the corresponding $\fb_*$ from some feature vectors $Z_*=\lbrace \zb_*^{(i)} \rbrace$ that are not in the training set.
To do so, we first identify the following joint distribution:
\begin{align}
\begin{bmatrix}
\yb \\
\fb_*
\end{bmatrix}
\sim 
\mathcal{N}\Bigg(\textbf{0},
\begin{bmatrix}
K + \sigma_n^2 I & K_* \\
K_*^\top & K_{**},
\end{bmatrix}
\Bigg),
\end{align}
where $K = K(Z, Z)$, $K_*= K(Z, Z_*)$, and $K_{**}=K(Z_*, Z_*)$ are covariance matrices and $K_{ij}(Z, Z') = k\left(\zb^{(i)},\zb'^{(j)}\right)$.
We then work out the conditional probability distribution of $\fb_*$ given $Z, \y, Z_*$ 
\begin{align}
    \fb_* | Z, \yb, Z_* \sim \mathcal{N}(\bar{\fb_*}, \Sigmab_*),
\end{align}
where 
\begin{align}
   \bar{\fb_*} &= K_*^\top (K + \sigma_n^2 I)^{-1}\yb, \nonumber \\
   \Sigmab_* &= K_{**} -  K_*^\top (K + \sigma_n^2 I)^{-1} K_*.
\end{align}
While $\bar{\fb}_*$ is our prediction for $\fb_*$, $\Sigmab_*$ can be used for constructing the confidence interval of the prediction.
Note that the prediction $\bar{\fb_*}$ depends on the training data $Z$ and $\yb$ explicitly, which is different from parametric models like MLP.
A schematic example for GPR is shown in Fig.~\ref{Fig:regression}(b) that contains the observed data, predicted values, and a 95$\%$ confidence interval of prediction.

These two classes of functions each have their own advantages and disadvantages. 
We will use both of them as our surrogate models for the edge update function in the following section and compare their performance in the problems that we are interested in here.

\section{Construction of the MGN based approach}
\label{Sec:construction}

In this section we will introduce the procedures of constructing the MGN based approach.
We start by identifying the physical constraints needed to be imposed over the edge update function.
Based on a proper form through which these constraints can be strictly imposed,
we train the edge update function to best approximate the mechanics of a given CMM at the cellular scale.
The performance of MLP and GPR as our surrogate for the edge update functions is also compared.

\subsection{Physical constraints}

\begin{figure}[!ht]
    \centering 
    \scalebox{0.9}{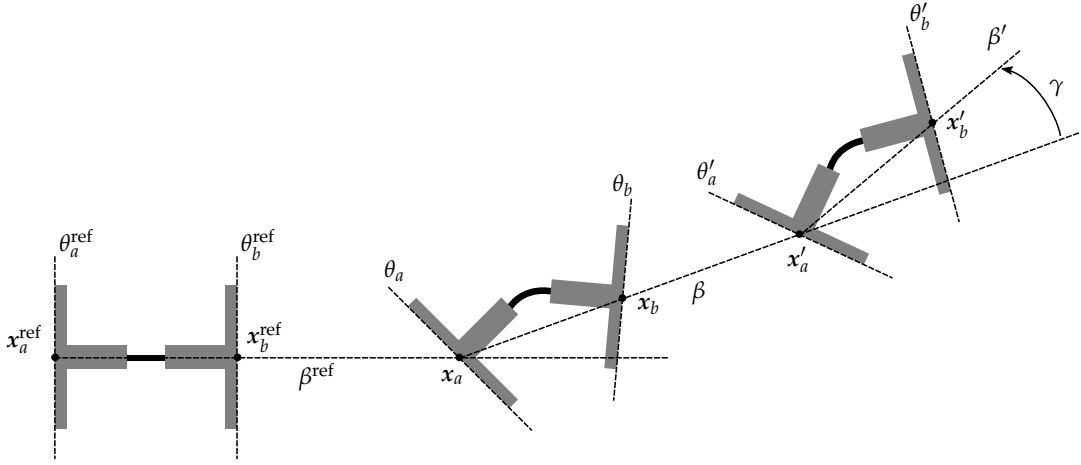}
    \caption{A building block of the cross-spring system in its reference configuration (leftmost), first spatial configuration (middle), and second spatial configuration (rightmost).
    The two spatial configurations differ only by a rigid body motion.}
    \label{Fig:objectivity}
\end{figure}

Before proceeding to the training process, we need to first discuss the constrains needed for the edge update function $\psi$.
Now, consider the cross-spring system in Fig.~\ref{Fig:objectivity}.  
In MGN, as defined in the previous section, we assume that $\psi$ is a function of both $\cb_a=(\x_a^{\textrm{ref}}, \theta_a^{\textrm{ref}}, \x_a, \theta_a)$ and $\cb_b=(\x_b^{\textrm{ref}}, \theta_b^{\textrm{ref}}, \x_b, \theta_b)$.
However, as $\psi$ represents the elastic energy of the corresponding unit-cell, it should be invariant under a global rigid body motion.
For example, the middle and right panels of Fig.~\ref{Fig:objectivity}.
These two configurations differ only by a translation and a rotation of angle $\gamma$ and therefore should have the same elastic energy, even though their nodal features are different. 


Suppose we denote the nodal features of crosses $a$ and $b$ in the rightmost panel as  $\cb_a'=(\x_a^{\textrm{ref}}, \theta_a^{\textrm{ref}}, \x_a', \theta_a')$ and $\cb_b'=(\x_b^{\textrm{ref}}, \theta_b^{\textrm{ref}}, \x_b', \theta_b')$, respectively.
Assume $\Qb \in SO(2)$ is an in-plane rotation of angle $\gamma$, which has the following matrix form:
\begin{align} \label{Eq:Q}
[\Qb] = 
\begin{bmatrix} 
\textrm{cos}(\gamma) & - \textrm{sin}(\gamma)  \\
\textrm{sin}(\gamma) & \textrm{cos}(\gamma)
\end{bmatrix},
\end{align}
then we know that the generalized coordinates between the two spatial configurations have the following relation:
\begin{align} \label{Eq:objectivity}
    \x_b' - \x_a' &= \Qb (\x_b - \x_a),  \nonumber \\
    \theta_a' &= \theta_a + \gamma,  \nonumber \\
    \theta_b' &= \theta_b + \gamma.
\end{align}
In order for the edge update function to be invariant under any in-plane rigid body motion, we propose the following form of energy $\tilde{\psi}$ such that
\begin{align} \label{Eq:energy_two}
    \tilde{\psi}(\tilde{\theta}_a, \tilde{\theta}_b, d) &= \psi(\cb_a, \cb_b), \nonumber \\
    \tilde{\theta}_a &= (\theta_a - \theta_a^{\textrm{ref}}) - (\beta - \beta^{\textrm{ref}}), \nonumber \\
    \tilde{\theta}_b &= (\theta_b - \theta_b^{\textrm{ref}}) - (\beta - \beta^{\textrm{ref}}), \nonumber \\
    d &= |\x_b - \x_a| - |\x_b^{\textrm{ref}} - \x_a^{\textrm{ref}}|,
\end{align}
where $\beta^{\textrm{ref}}=\textrm{angle}(\x_b^{\textrm{ref}}-\x_a^{\textrm{ref}})$ and $\beta=\textrm{angle}(\x_b - \x_a)$ (see Fig.~\ref{Fig:objectivity}) are the polar angles of the spring in the reference and spatial configurations, respectively.
Similarly, we have the following quantities for the second spatial configuration
\begin{align}
    \tilde{\psi}(\tilde{\theta}_a', \tilde{\theta}_b', d') &= \psi(\cb_a', \cb_b'), \nonumber \\
    \tilde{\theta}_a' &= (\theta_a' - \theta_a^{\textrm{ref}}) - (\beta' - \beta^{\textrm{ref}}), \nonumber \\
    \tilde{\theta}_b' &= (\theta_b' - \theta_b^{\textrm{ref}}) - (\beta' - \beta^{\textrm{ref}}), \nonumber \\
    d' &= |\x_b' - \x_a'| - |\x_b^{\textrm{ref}} - \x_a^{\textrm{ref}}|,
\end{align}

Obviously, $\tilde{\theta}_a=\tilde{\theta}_a'$, $\tilde{\theta}_b=\tilde{\theta}_b'$, and $d=d'$, under any rigid body motion.
Then, if we use $\tilde{\psi}$, then the energy is invariant in the two spatial configurations, namely, we have $\tilde{\psi}(\tilde{\theta}_a, \tilde{\theta}_b, d)=\tilde{\psi}(\tilde{\theta}_a', \tilde{\theta}_b', d')$.
The original potential energy function $\psi$ takes $\x_a$, $\theta_a$, $\x_b$, $\theta_b$ as its input (a six dimensional vector), while $\tilde{\psi}$ only takes $\tilde{\theta_a}$, $\tilde{\theta_b}$, $d$ as the input (a three dimensional vector).
By imposing the constraint in this way, the number of degrees of freedom are also reduced.

With this $\tilde{\psi}$ on hand, computing $\Psi$ for given $(\Xi, \Theta)$ is straightforward and in principle, also straightforward for determining $\frac{\partial \Psi}{\partial \Xi}$ and $\frac{\partial \Psi}{\partial \Theta}$.
While this can be carried out using chain rule, the calculation can be highly laborious and error-prone.
In this work, we make use of the open-source package \texttt{JAX}~\citep{jax2018github} to automate this procedure by using automatic differentiation.
In \texttt{JAX}, we write $\Psi$ as a function of $(\Xi, \Theta)$ in \texttt{Python}, and a single call of \texttt{jax.grad} returns the corresponding derivative function that we need. 
Interested readers are referred to our code on github for details. 

%
%
%

\subsection{Training, validation, and test}

Training, validation, and testing are performed under the framework of standard supervised learning~\citep{bishop2006pattern}.
Through the data generation process introduced in \ref{App:data_prep}, we obtain $D=\lbrace (\zb;y)^{(i)}\rbrace_{i=1:|D|}$ ($|D|=1000$) for each of the five pore shapes. 
The label $y$ is obtained by finite element simulation of the corresponding building block under a deformation that is determined by the feature $\zb$ (see Fig.~\ref{Fig:beam}).
Separate models are trained over these five data sets for explicit comparison of model performance between different shapes.

We split each data set for training, validation, and test with an 80$\%$-10$\%$-10$\%$ ratio.
As mentioned earlier, both MLP and GPR are used as our surrogate for comparison.
Models are trained over the training data set, their performance compared over the validation data set, and only the model with the best performance is tested on the test data set.

For MLPs, we choose the mean squared error (MSE) as our loss function,  and use mini-batch stochastic gradient descent with Adam optimizer~\citep{kingma2014adam}.
Here, we choose three different MLPs (labelled as MLP $\#$1, MLP $\#$2, and MLP $\#$3) with increasing model complexity and compare their performance.
While it is almost impossible to exhaustively search for the best set of hyperparameters, the selected three MLPs are representative and their hyperparameters are listed in Table~\ref{Tab:MLP}.

\begin{table}[t]
\centering
\scriptsize
\begin{tabular}{@{}cccccc@{}}
\toprule
\multicolumn{1}{c}{} & \multicolumn{1}{c}{Hidden layers} & \multicolumn{1}{c}{Layer width} & \multicolumn{1}{c}{Learning rate} & \multicolumn{1}{c}{Batch size} & \multicolumn{1}{c}{Activation function} \\ \midrule
MLP \#1              & 2                                 & 32                              & $4\times 10^{-4}$                 & 32                             & ReLU                                    \\ \midrule
MLP \#2              & 4                                 & 64                              & $2\times 10^{-4}$                 & 32                             & ReLU                                    \\ \midrule
MLP \#3              & 8                                 & 128                             & $1\times 10^{-1}$                 & 32                             & ReLU                                    \\ \bottomrule
\end{tabular}
\caption{Hyperparameters of the three MLPs used in this work. }
\label{Tab:MLP}
\end{table}
 
\begin{table}[H]
\centering
\scriptsize
\begin{tabular}{cccccc}
\hline
             & shape A           & shape B           & shape C           & shape D           & shape E           \\ \hline
$\sigma^2$   & 0.672            & 0.751            & 0.796            & 0.792            & 0.759            \\ \hline
$l$          & 0.890            & 0.874            & 0.865            & 0.866            & 0.872            \\ \hline
$\sigma^2_n$ & $5\times10^{-5}$ & $5\times10^{-5}$ & $5\times10^{-5}$ & $5\times10^{-5}$ & $5\times10^{-5}$ \\ \hline
\end{tabular}
\caption{Optimized hyperparameters for GPR for different pore shapes.}
\label{Tab:GPR}
\end{table}

For GPR, we determine the set of optimal hyperparameters $(\sigma^2, l, \sigma^2_n)$ based on the training data set by maximizing the log marginal likelihood~\citep{rasmussen2003gaussian} with quasi-Newton methods like L-BFGS-B~\citep{byrd1995limited,zhu1997algorithm}.
The optimized hyperparameters are reported in Table~\ref{Tab:GPR}.

\begin{figure}[t]
    \centering 
    \scalebox{0.95}{
\begingroup%
  \makeatletter%
  \providecommand\color[2][]{%
    \errmessage{(Inkscape) Color is used for the text in Inkscape, but the package 'color.sty' is not loaded}%
    \renewcommand\color[2][]{}%
  }%
  \providecommand\transparent[1]{%
    \errmessage{(Inkscape) Transparency is used (non-zero) for the text in Inkscape, but the package 'transparent.sty' is not loaded}%
    \renewcommand\transparent[1]{}%
  }%
  \providecommand\rotatebox[2]{#2}%
  \newcommand*\fsize{\dimexpr\f@size pt\relax}%
  \newcommand*\lineheight[1]{\fontsize{\fsize}{#1\fsize}\selectfont}%
  \ifx\svgwidth\undefined%
    \setlength{\unitlength}{503.17101574bp}%
    \ifx\svgscale\undefined%
      \relax%
    \else%
      \setlength{\unitlength}{\unitlength * \real{\svgscale}}%
    \fi%
  \else%
    \setlength{\unitlength}{\svgwidth}%
  \fi%
  \global\let\svgwidth\undefined%
  \global\let\svgscale\undefined%
  \makeatother%
  \begin{picture}(1,0.52112066)%
    \lineheight{1}%
    \setlength\tabcolsep{0pt}%
    \put(0.15684789,0.28372656){\color[rgb]{0,0,0}\makebox(0,0)[lt]{\lineheight{1.25}\smash{\begin{tabular}[t]{l}(a)\end{tabular}}}}%
    \put(0.50222183,0.28372656){\color[rgb]{0,0,0}\makebox(0,0)[lt]{\lineheight{1.25}\smash{\begin{tabular}[t]{l}(b)\end{tabular}}}}%
    \put(0.85041658,0.28372656){\color[rgb]{0,0,0}\makebox(0,0)[lt]{\lineheight{1.25}\smash{\begin{tabular}[t]{l}(c)\end{tabular}}}}%
    \put(0.32326779,0.0027841){\color[rgb]{0,0,0}\makebox(0,0)[lt]{\lineheight{1.25}\smash{\begin{tabular}[t]{l}(d)\end{tabular}}}}%
    \put(0.69661771,0.0027841){\color[rgb]{0,0,0}\makebox(0,0)[lt]{\lineheight{1.25}\smash{\begin{tabular}[t]{l}(e)\end{tabular}}}}%
    \put(0,0){\includegraphics[width=\unitlength,page=1]{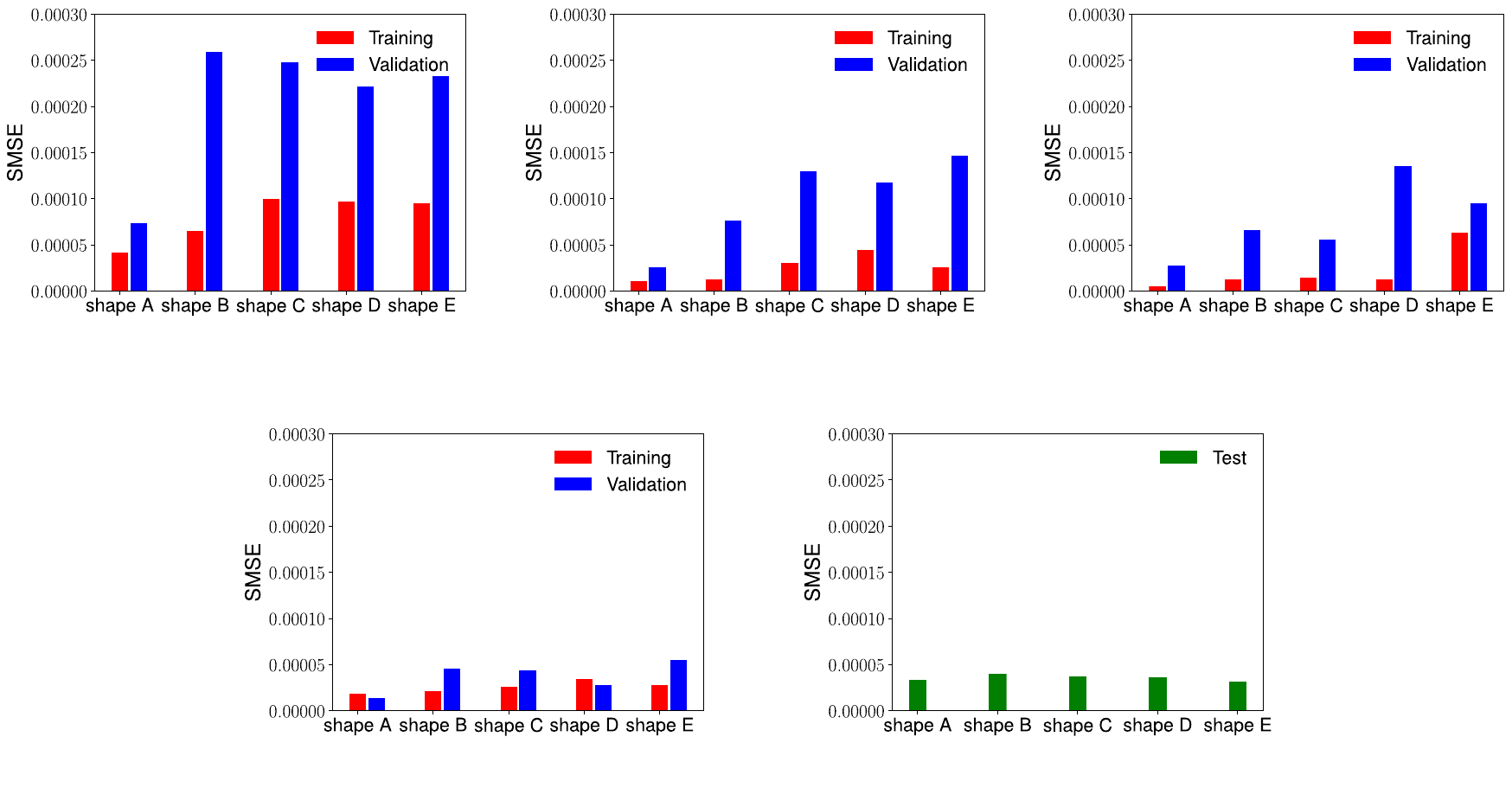}}%
  \end{picture}%
\endgroup%
}
    \caption{Training and validation errors of MLP\#1 in (a), MLP\#2 in (b), MLP\#3 in (c), and for GPR in (d).
    Only the test errors of GPR are reported in (e) since it is the optimal model.}
    \label{Fig:train_val_test}
\end{figure}
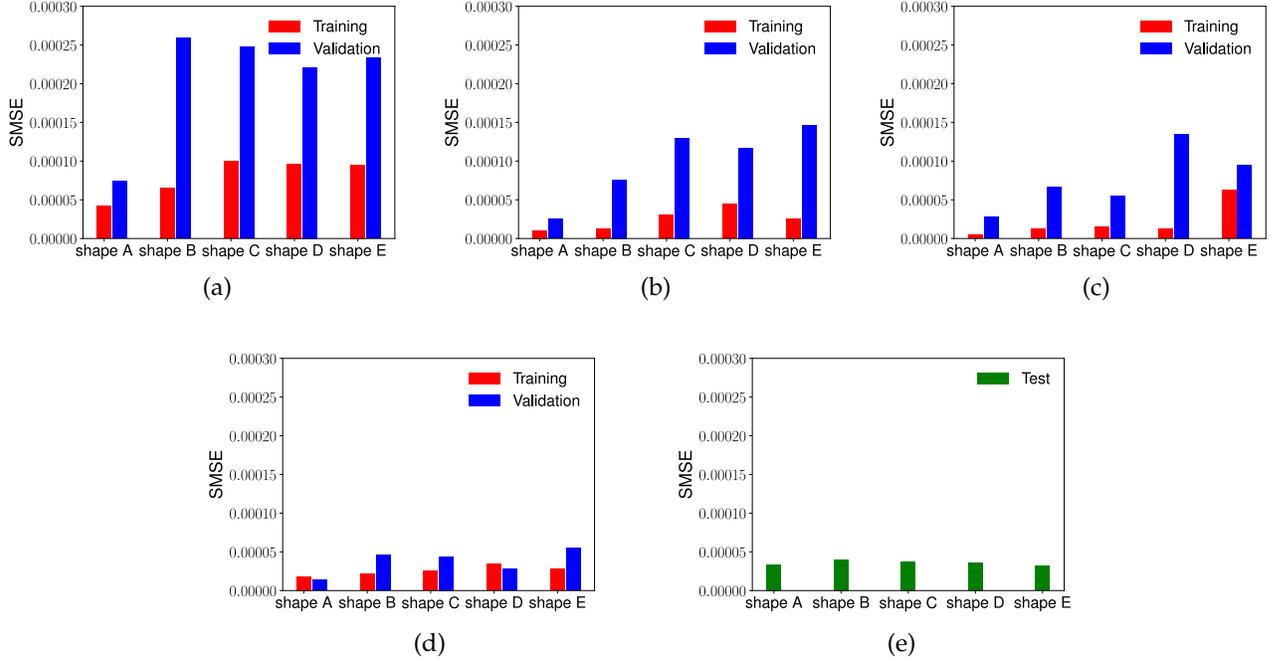

For the three MLP models and one GPR model listed above, we use the scaled mean square error (SMSE) as the criterion to choose the optimal model, which is defined as
\begin{align}
    \textrm{SMSE} = \frac{1}{N_s} \sum_{i=1}^{N_{s}} (\bar{y}_{\textrm{true}} - \bar{y}_{\textrm{pred}})^2 \quad \textrm{with} \quad \bar{y} := \frac{y - y_{\textrm{min}}}{y_{\textrm{max}} - y_{\textrm{min}}},
\end{align}
where $N_s$ is the number of samples considered, $\bar{y}_{\textrm{true}}$ is the scaled true output, $\bar{y}_{\textrm{pred}}$ is the scaled predicted output, and  $(y_{\textrm{min}}, y_{\textrm{max}})$ are the lower and upper bounds for min-max scaling.
We report training and validation SMSEs for the three MLPs and GPR in Fig.~\ref{Fig:train_val_test}(a), (b), (c), and (d), respectively. 
With increasing model complexity, MLPs show lower training errors.
However, all three MLPs do not generalize well to the validation data set and are prone to overfitting.
In contrast, GPR shows relatively low errors on both training and validation data sets, which indicates that GPR outperforms MLPs for our problem, and this observation is consistent for all five shapes.
Therefore, we select GPR as our best model and will exclusively use GPR in all following discussions.
The test error of GPR is reported in Fig.~\ref{Fig:train_val_test}(e), less than $5\times10^{-4}$, which confirms good generalizability of the trained model.
For this reason, we chose \textbf{GPR} as our surrogate model for the edge update functions.
 
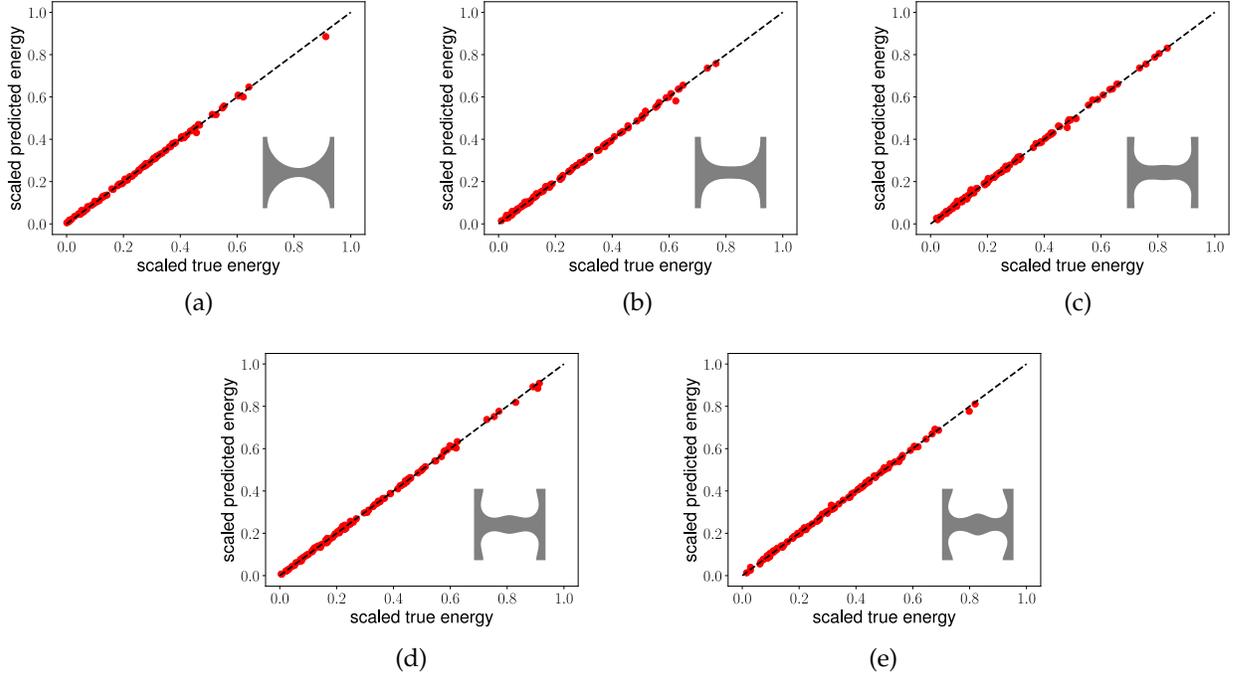
\begin{figure}[!ht]
    \centering 
    \scalebox{0.95}{
\begingroup%
  \makeatletter%
  \providecommand\color[2][]{%
    \errmessage{(Inkscape) Color is used for the text in Inkscape, but the package 'color.sty' is not loaded}%
    \renewcommand\color[2][]{}%
  }%
  \providecommand\transparent[1]{%
    \errmessage{(Inkscape) Transparency is used (non-zero) for the text in Inkscape, but the package 'transparent.sty' is not loaded}%
    \renewcommand\transparent[1]{}%
  }%
  \providecommand\rotatebox[2]{#2}%
  \newcommand*\fsize{\dimexpr\f@size pt\relax}%
  \newcommand*\lineheight[1]{\fontsize{\fsize}{#1\fsize}\selectfont}%
  \ifx\svgwidth\undefined%
    \setlength{\unitlength}{489.33469109bp}%
    \ifx\svgscale\undefined%
      \relax%
    \else%
      \setlength{\unitlength}{\unitlength * \real{\svgscale}}%
    \fi%
  \else%
    \setlength{\unitlength}{\svgwidth}%
  \fi%
  \global\let\svgwidth\undefined%
  \global\let\svgscale\undefined%
  \makeatother%
  \begin{picture}(1,0.54671866)%
    \lineheight{1}%
    \setlength\tabcolsep{0pt}%
    \put(0,0){\includegraphics[width=\unitlength,page=1]{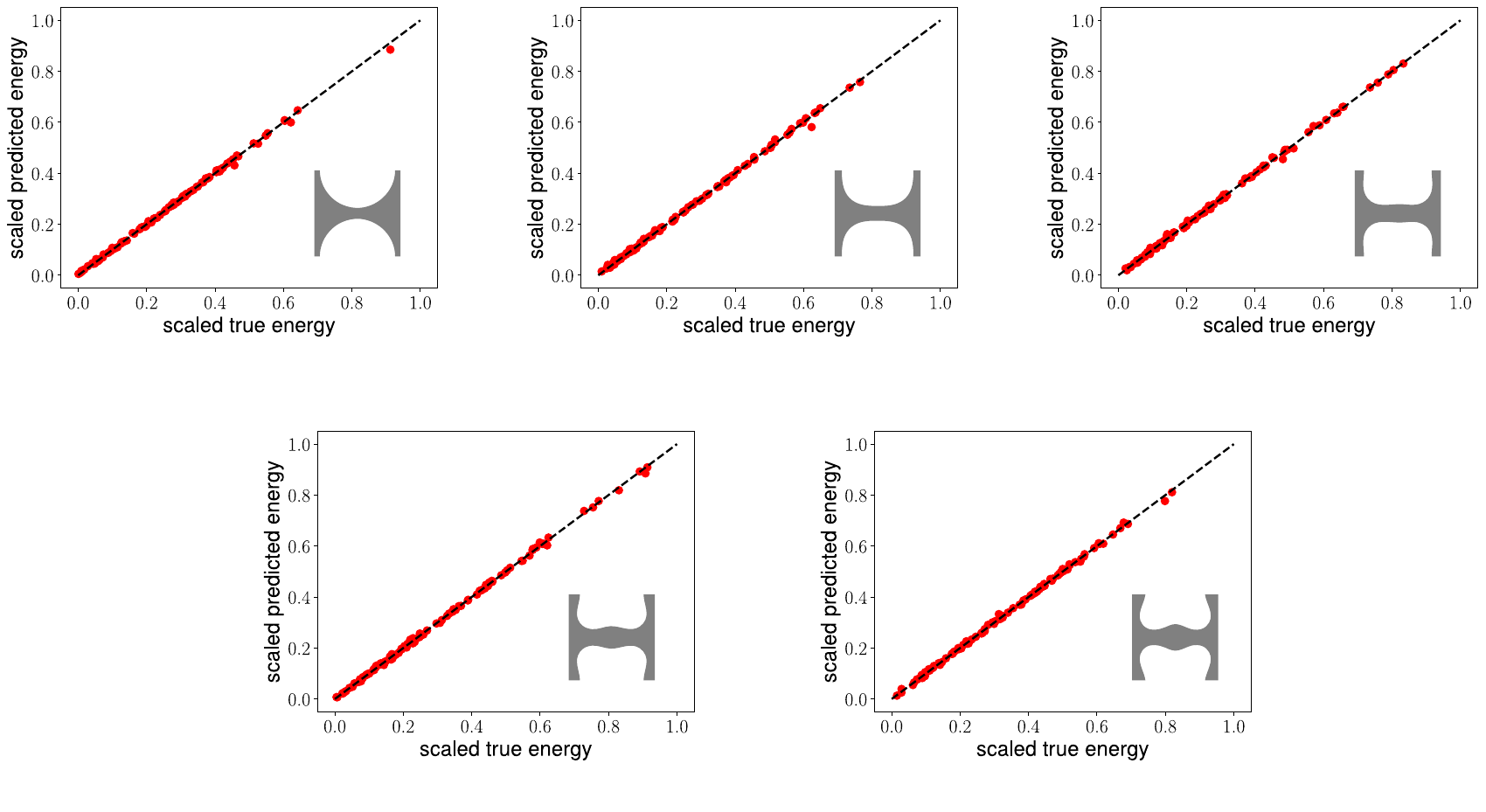}}%
    \put(0.14806332,0.2917491){\color[rgb]{0,0,0}\makebox(0,0)[lt]{\lineheight{1.25}\smash{\begin{tabular}[t]{l}(a)\end{tabular}}}}%
    \put(0.50320297,0.2917491){\color[rgb]{0,0,0}\makebox(0,0)[lt]{\lineheight{1.25}\smash{\begin{tabular}[t]{l}(b)\end{tabular}}}}%
    \put(0.86124324,0.2917491){\color[rgb]{0,0,0}\makebox(0,0)[lt]{\lineheight{1.25}\smash{\begin{tabular}[t]{l}(c)\end{tabular}}}}%
    \put(0.3191889,0.00286282){\color[rgb]{0,0,0}\makebox(0,0)[lt]{\lineheight{1.25}\smash{\begin{tabular}[t]{l}(d)\end{tabular}}}}%
    \put(0.70309559,0.00286282){\color[rgb]{0,0,0}\makebox(0,0)[lt]{\lineheight{1.25}\smash{\begin{tabular}[t]{l}(e)\end{tabular}}}}%
  \end{picture}%
\endgroup%
}
    \caption{Comparison between the true output and the prediction using the trained GPR model for the five shapes.}
    \label{Fig:pred_true}
\end{figure}

To further study the performance of the trained GPR model, we plot the scaled true energy $\bar{y}_{\textrm{true}}$ versus the scaled predicted energy $\bar{y}_{\textrm{pred}}$.
For consistency, the scaling uses training data as the reference.
We show the results in Fig~\ref{Fig:pred_true}.
The red data points mostly are aligned with the dashed reference line, which shows a good agreement between the true energy and the prediction.

\begin{figure}[!ht]
    \centering 
    \scalebox{1.}{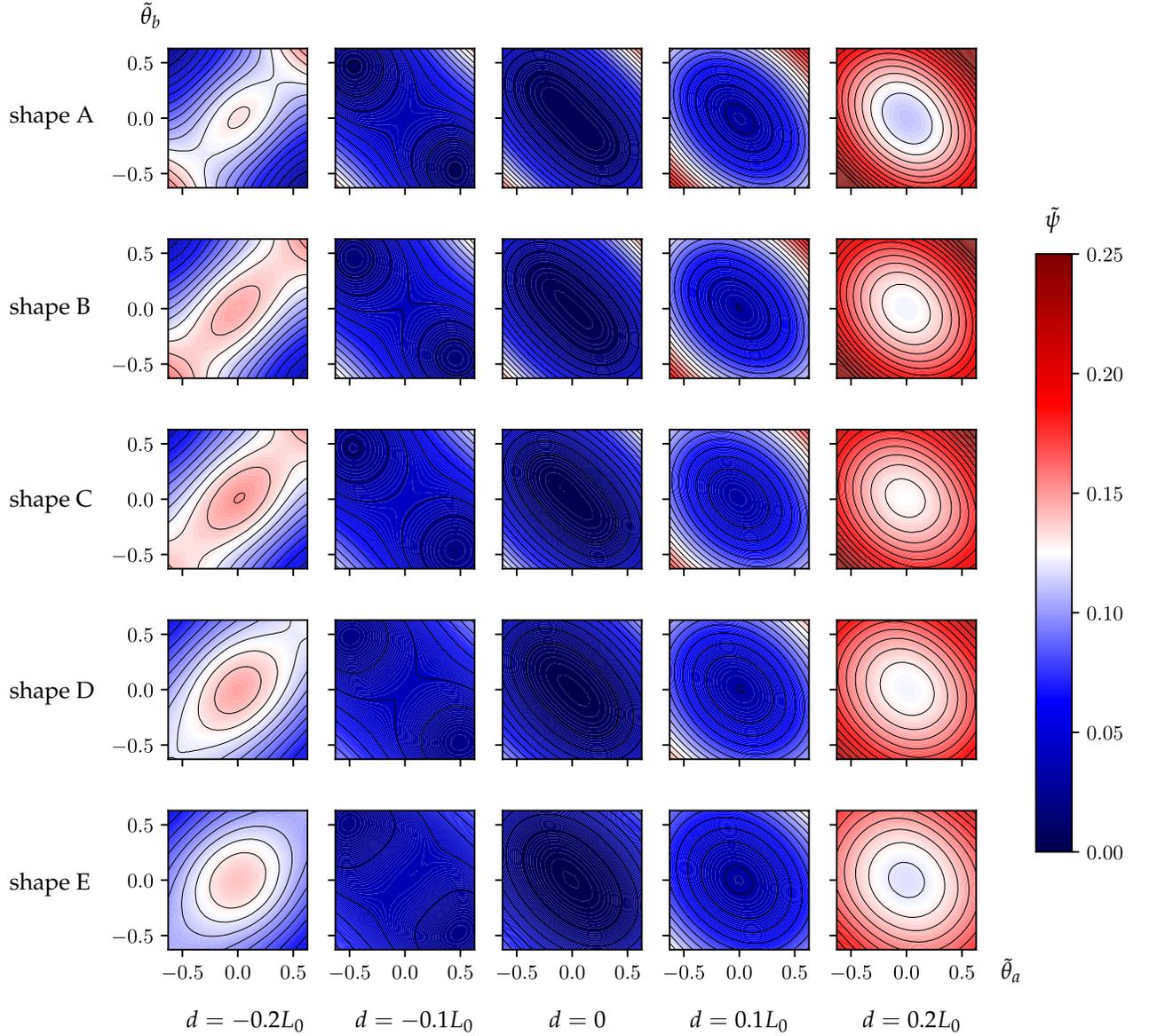}
    \caption{Energy contours for the five shapes. The plots are generated based on the trained GPR model.}
    \label{Fig:contour}
\end{figure}

The trained GPR model is an approximate surrogate for the energy function $\tilde{\psi}(\tilde{\theta}_a, \tilde{\theta}_b, d)$.
In Fig.~\ref{Fig:contour}, we plot the energy contours for the five shapes.
For each shape, we vary $d$ from $-0.2L_0$ to $0.2L_0$, which corresponds to the spring from being compressed to being stretched, and plot $\tilde{\theta}_a$ v.s. $\tilde{\theta}_b$ in the range of $(-\frac{\pi}{5}, \frac{\pi}{5})\times(-\frac{\pi}{5}, \frac{\pi}{5})$.
We observe from the contour plots that the building blocks with five different shapes show similar responses when $d>0$.
This indicates that under tensile forces, the springs generally have similar mechanical behaviors, which is consistent with our previous work~\citep{xue2020data}.
On the contrary, the energy contours show richer responses that are highly nonlinear when $d<0$, i.e., under compression.
For example, for the building block with shape A under $d=0.2L_0$, only one minimum is observed on the contour plot, whereas when $d=-0.2L_0$, the energy landscape is clearly more complex and it has two minima ($\tilde{\theta}_a=\frac{\pi}{5},  \tilde{\theta}_b=-\frac{\pi}{5})$ and ($\tilde{\theta}_a=-\frac{\pi}{5},  \tilde{\theta}_b=\frac{\pi}{5})$.

\newpage

\section{Numerical examples}
\label{Sec:numerical}

In this section, we use the trained GPR model as the edge update function of our MGN to study the mechanics of CMMs via several representative numerical examples.
The first example is when the CMM is under quasi-static loading, which serves as benchmarks for our MGN based approach.
In the second example, we study how an impulse propagates through the CMM. 
In these two examples, DNS is still possible and its performance is compared against our MGN based approach.
We then use the MGN based approach to study the dynamic response of CMMs with larger sizes, of which the DNS is computationally too expensive.


\subsection{Quasi-static loading}

Previous studies have shown that the mechanical properties of CMMs are highly sensitive to the pore shapes~\citep{bertoldi2010negative,overvelde2012compaction, overvelde2014relating,xue2020data}.
Under uniaxial compression, mechanical instabilities at the scale of unit-cell often lead to pattern transformations at the structural scale. 
As a benchmark, we employ MGN based approach to study two representative CMMs, one with negative Poisson ratio, and the other with bifurcation at the unit-cell scale.
The results obtained from MGN based approach are compared with those from DNS.
To minimize the inertial effect, DNS is carried out in a quasi-static manner, while a large damping is added to Eq.~(\ref{Eq:strong_cs}).

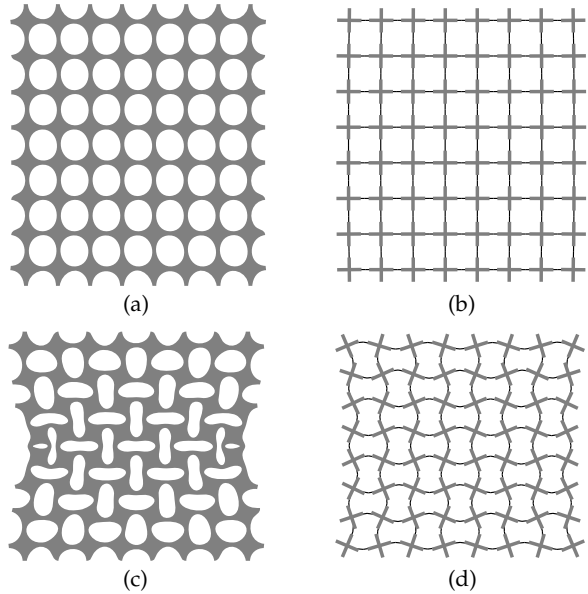
\begin{figure}[!ht]
    \centering 
    \scalebox{0.85}{
\begingroup%
  \makeatletter%
  \providecommand\color[2][]{%
    \errmessage{(Inkscape) Color is used for the text in Inkscape, but the package 'color.sty' is not loaded}%
    \renewcommand\color[2][]{}%
  }%
  \providecommand\transparent[1]{%
    \errmessage{(Inkscape) Transparency is used (non-zero) for the text in Inkscape, but the package 'transparent.sty' is not loaded}%
    \renewcommand\transparent[1]{}%
  }%
  \providecommand\rotatebox[2]{#2}%
  \newcommand*\fsize{\dimexpr\f@size pt\relax}%
  \newcommand*\lineheight[1]{\fontsize{\fsize}{#1\fsize}\selectfont}%
  \ifx\svgwidth\undefined%
    \setlength{\unitlength}{278.02855383bp}%
    \ifx\svgscale\undefined%
      \relax%
    \else%
      \setlength{\unitlength}{\unitlength * \real{\svgscale}}%
    \fi%
  \else%
    \setlength{\unitlength}{\svgwidth}%
  \fi%
  \global\let\svgwidth\undefined%
  \global\let\svgscale\undefined%
  \makeatother%
  \begin{picture}(1,0.99632217)%
    \lineheight{1}%
    \setlength\tabcolsep{0pt}%
    \put(0,0){\includegraphics[width=\unitlength,page=1]{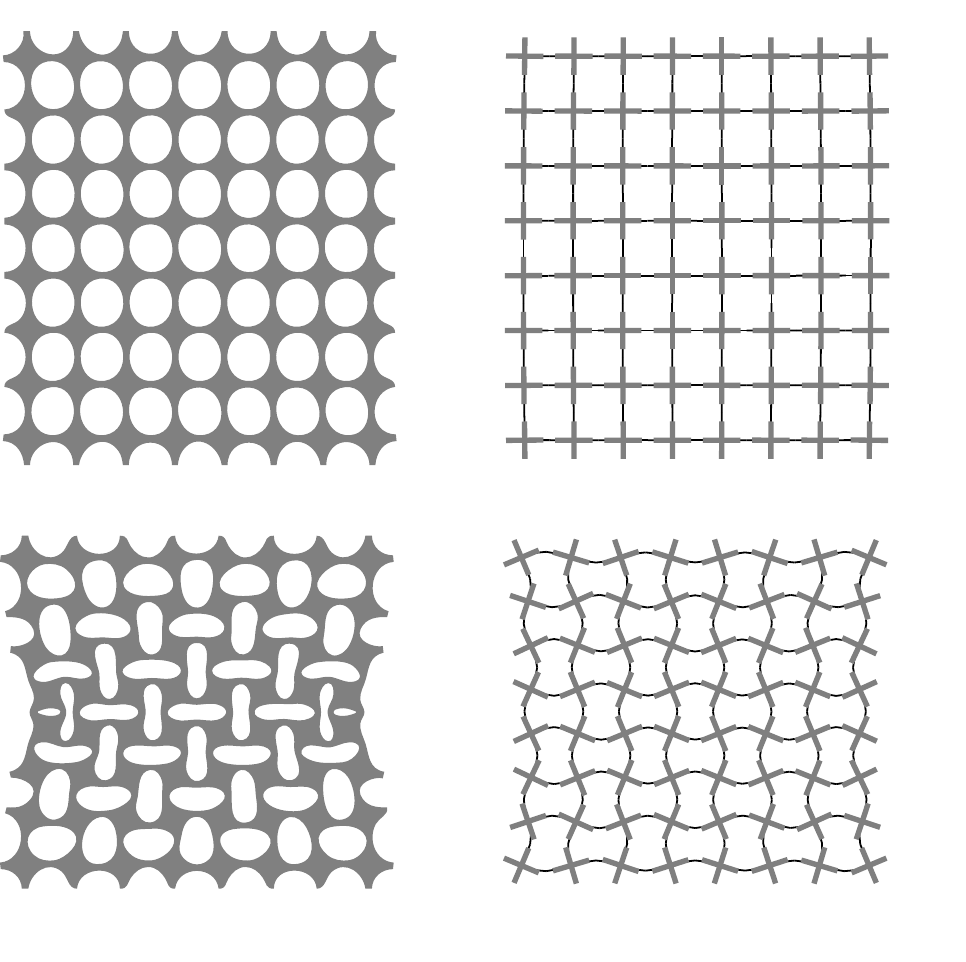}}%
    \put(0.18280853,0.47498146){\color[rgb]{0,0,0}\makebox(0,0)[lt]{\lineheight{1.25}\smash{\begin{tabular}[t]{l}(a)\end{tabular}}}}%
    \put(0.70112629,0.47498146){\color[rgb]{0,0,0}\makebox(0,0)[lt]{\lineheight{1.25}\smash{\begin{tabular}[t]{l}(b)\end{tabular}}}}%
    \put(0.18280853,0.03470044){\color[rgb]{0,0,0}\makebox(0,0)[lt]{\lineheight{1.25}\smash{\begin{tabular}[t]{l}(c)\end{tabular}}}}%
    \put(0.70112629,0.03470044){\color[rgb]{0,0,0}\makebox(0,0)[lt]{\lineheight{1.25}\smash{\begin{tabular}[t]{l}(d)\end{tabular}}}}%
  \end{picture}%
\endgroup%
}
    \caption{Uniaxial tensile and compression tests for structures with shape A.  
    (a) and (c) show the final configurations of an 8$\times$8 continuum CMM computed by DNS.
    (b) and (d) show the configurations of the corresponding cross-spring systems computed by MGN simulation.}
    \label{Fig:statics_poreA}
\end{figure}

The first example in this subsection studies a CMM with perfect circular pores which becomes auxetic beyond certain critical compression.
We conduct DNS for this kind of CMMs made of an $8\times8$ array of unit-cells under uniaxial loading.
The deformed shapes are shown in Fig.~\ref{Fig:statics_poreA}(a) for $10\%$ tension and Fig.~\ref{Fig:statics_poreA}(c) for $10\%$ compression.
It is expected that tensile loadings do not trigger instabilities.
In contrast, compressive loadings lead to a clear pattern transformation such that the circular pores are deformed into alternatively arranged ellipses.
On the structural scale, the structure tends to shrink in the horizontal direction, indicating a negative Poisson's ratio.
Under the same external loading conditions, we perform MGN based simulation for a cross-spring system with the same size.
The edge update function used in this MGN is chosen to be the one trained for shape A (circular pore).
As shown in Fig.~\ref{Fig:statics_poreA}(a) and (b), results produced by the MGN based approach agree well with those by DNS, except for some parts on the boundary when the CMM is under compression.

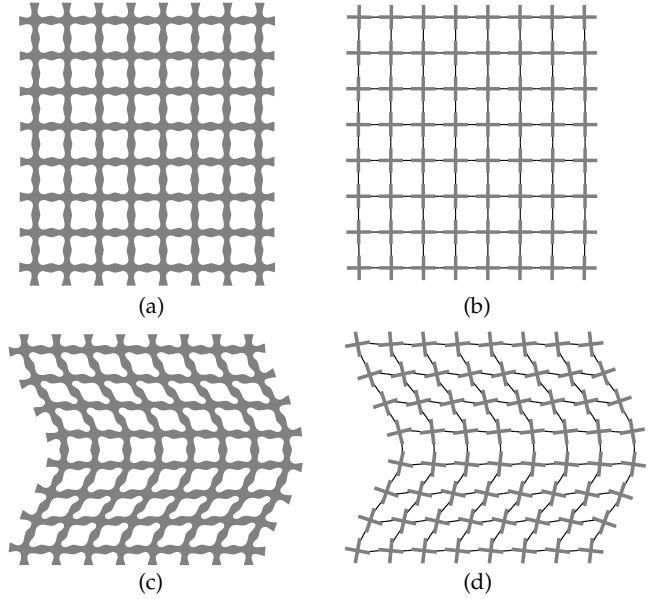
\begin{figure}[!ht]
    \centering 
    \scalebox{0.85}{
\begingroup%
  \makeatletter%
  \providecommand\color[2][]{%
    \errmessage{(Inkscape) Color is used for the text in Inkscape, but the package 'color.sty' is not loaded}%
    \renewcommand\color[2][]{}%
  }%
  \providecommand\transparent[1]{%
    \errmessage{(Inkscape) Transparency is used (non-zero) for the text in Inkscape, but the package 'transparent.sty' is not loaded}%
    \renewcommand\transparent[1]{}%
  }%
  \providecommand\rotatebox[2]{#2}%
  \newcommand*\fsize{\dimexpr\f@size pt\relax}%
  \newcommand*\lineheight[1]{\fontsize{\fsize}{#1\fsize}\selectfont}%
  \ifx\svgwidth\undefined%
    \setlength{\unitlength}{286.78424334bp}%
    \ifx\svgscale\undefined%
      \relax%
    \else%
      \setlength{\unitlength}{\unitlength * \real{\svgscale}}%
    \fi%
  \else%
    \setlength{\unitlength}{\svgwidth}%
  \fi%
  \global\let\svgwidth\undefined%
  \global\let\svgscale\undefined%
  \makeatother%
  \begin{picture}(1,0.98211586)%
    \lineheight{1}%
    \setlength\tabcolsep{0pt}%
    \put(0,0){\includegraphics[width=\unitlength,page=1]{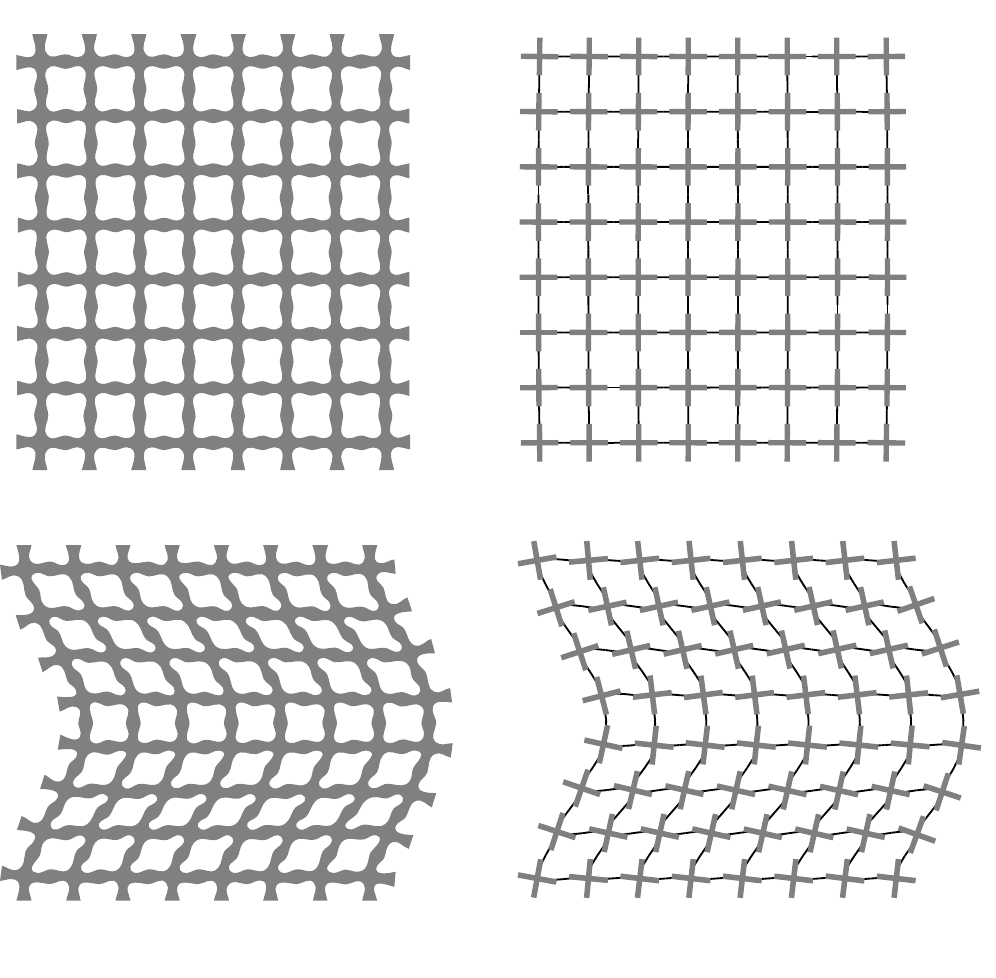}}%
    \put(0.20074081,0.46690001){\color[rgb]{0,0,0}\makebox(0,0)[lt]{\lineheight{1.25}\smash{\begin{tabular}[t]{l}(a)\end{tabular}}}}%
    \put(0.70323399,0.46690001){\color[rgb]{0,0,0}\makebox(0,0)[lt]{\lineheight{1.25}\smash{\begin{tabular}[t]{l}(b)\end{tabular}}}}%
    \put(0.20074081,0.04006106){\color[rgb]{0,0,0}\makebox(0,0)[lt]{\lineheight{1.25}\smash{\begin{tabular}[t]{l}(c)\end{tabular}}}}%
    \put(0.70323399,0.04006106){\color[rgb]{0,0,0}\makebox(0,0)[lt]{\lineheight{1.25}\smash{\begin{tabular}[t]{l}(d)\end{tabular}}}}%
  \end{picture}%
\endgroup%
}
    \caption{Uniaxial tensile and compression tests for structures with shape A.  
    (a) and (c) show the final static configurations of an 8$\times$8 continuum CMM computed by DNS.
    (b) and (d) show the configurations of the corresponding cross-spring systems computed by MGN simulation.}
    \label{Fig:statics_poreE}
\end{figure}

The second benchmark studies the CMM with pores of shape E.
It is known that such CMMs will start to bifurcate and have multiple stable states once the compression is beyond a threshold.
We conduct DNS for this kind of CMMs under uniaxial loadings, and show the results in Fig.~\ref{Fig:statics_poreE}(a) for $10\%$ tension and Fig.~\ref{Fig:statics_poreE}(c) for $10\%$ compression.
As shown, the structure buckles to one side under compression.
The same phenomenon is captured by MGN based simulation as shown in Fig.~\ref{Fig:statics_poreE}(d).

These two benchmarks show that MGN based simulation is able to produce results that agree well with DNS under quasi-static loadings.

\subsection{Impulse propagation}

\begin{figure}[!ht]
    \centering 
    \scalebox{0.95}{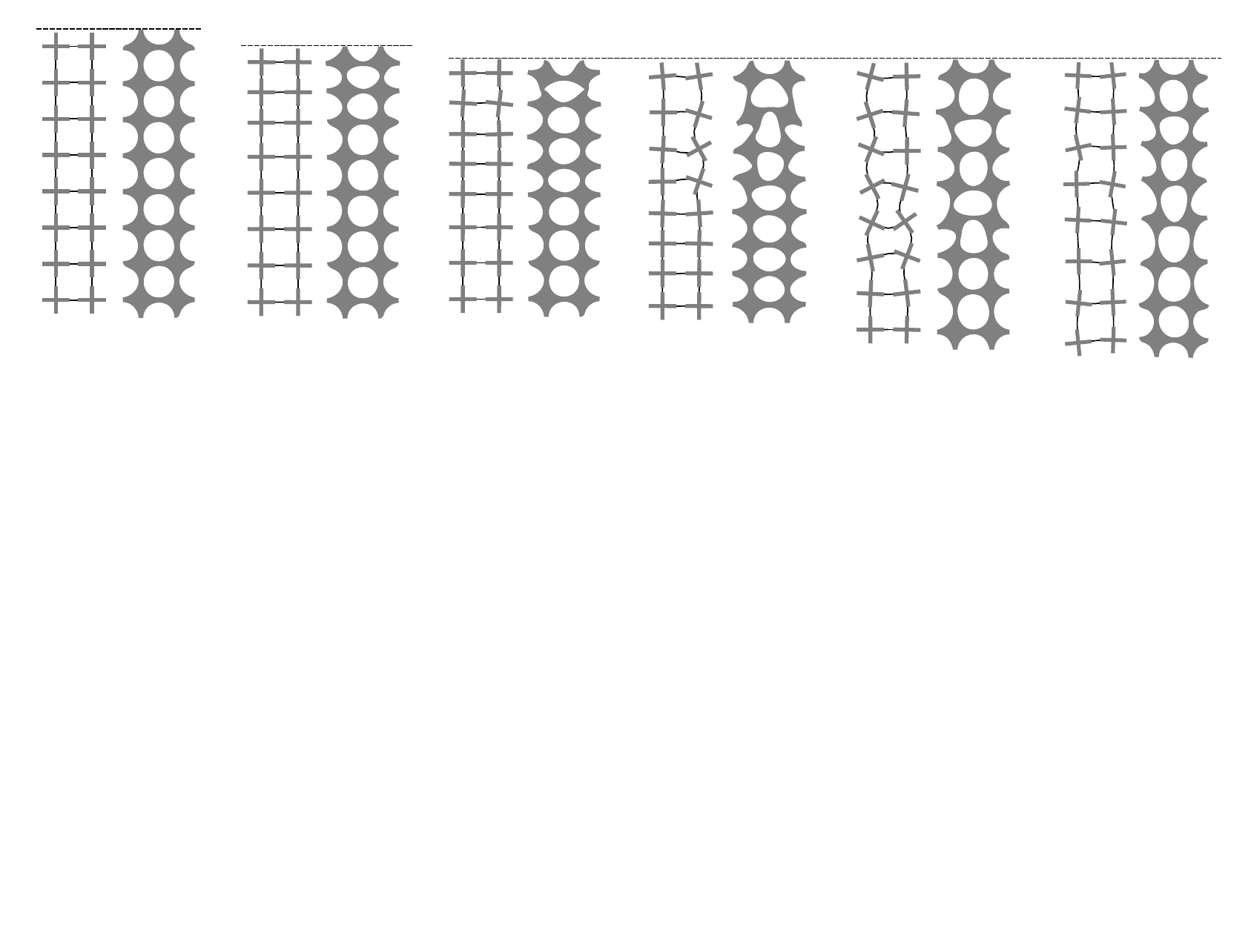}
    \caption{Longitudinal wave propagation in 2$\times$8 structures for both DNS and MGN results.
    Comparison of configurations at different time are shown in (a).
    Kinetic energy stored in the bottom row of units are recorded and shown in (b) for the cross-spring system, and in (c) for the continuum CMM.
    }
    \label{Fig:P_wave}
\end{figure}

We proceed to study the propagation of a impulse through CMMs.
As is shown in Fig.~\ref{Fig:P_wave}(a), we consider a CMM made of a 2$\times$8 array of unit-cells, which is at rest initially and subjected to an impulse at the top boundary.
We prescribe the following boundary condition on the upper side of the structure to approximate an impulse:
\begin{equation}
  U_2(X_1, 8L_0, t) =  \left\{
        \begin{array}{ll}
            -80L_0 t & \quad 0 \leq t \leq 0.01 \\
            -0.8L_0 & \quad t > 0.01
        \end{array}
    \right.
\end{equation}
All other parts of the boundary are traction free.

The corresponding cross-spring system is shown to the left of the continuum CMM.
The results of the continuum CMM are obtained via DNS and those of the cross-spring system from the MGN based simulation.
An interesting observation is that, when at $t=0.015$s, both DNS and MGN based approach suggest that symmetry breaking can happen for the wave profiles.

To study wave propagation through the structure, we track the kinetic energy stored in unit cell at the bottom of the structure (see the red box in Fig.~\ref{Fig:P_wave}(a)).
Before the wave reaches the bottom side, those units remain at rest and the kinetic energy is zero. 
Once the wave propagates to the bottom side, there is a sharp rise in its kinetic energy.
We plot the kinetic energy of the bottom unit-cell for all five shapes in Fig.~\ref{Fig:P_wave}(b) and (c), which are obtained from MGN based simulation and DNS respectively.
From the figure, we can observe that the results produced by MGN simulation are in good agreement with DNS.

As is observed in Fig.~\ref{Fig:P_wave}(b) and (c), both MGN and DNS results show that the times when the kinetic energy of the bottom row reaches the peak are almost identical for different pore shapes (except for pore shape E obtained from DNS).
In this particular example, such observation seems to suggest that the wave speed is quite insensitive to the pore shapes. 

\subsection{Computational performance }

\begin{figure}[H]
    \centering 
    \includegraphics[width=0.6\textwidth]{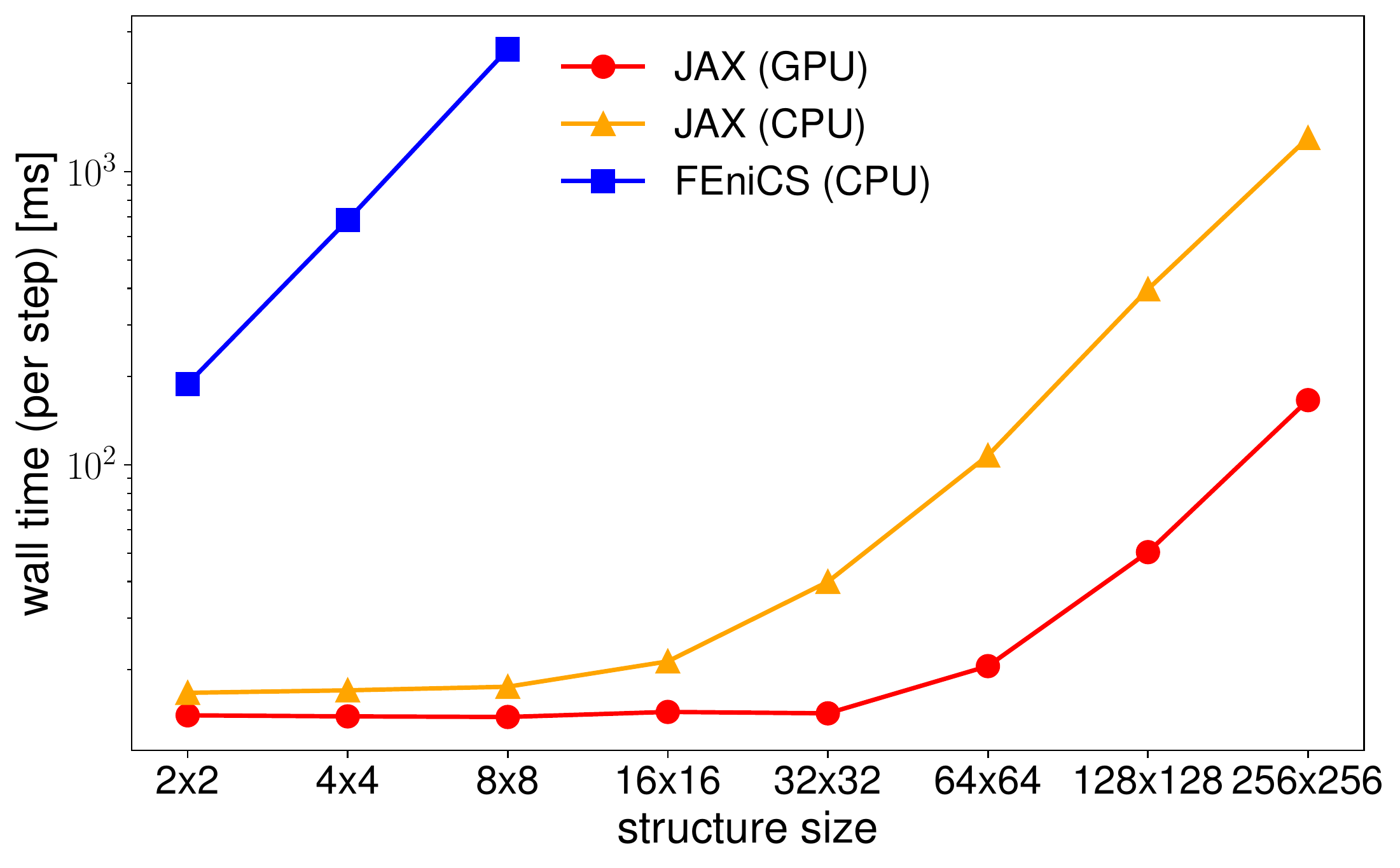}
    \caption{Comparison of computational performance between DNS with \texttt{FEniCS} (CPU), MGN with \texttt{JAX} (CPU), and MGN with \texttt{JAX} (GPU).
    The wall-clock time is for each integration step.}
    \label{Fig:performance}
\end{figure}

Previous numerical examples show that MGN simulations of cross-spring systems are accurate surrogates for the elastrodynamics of continuum CMMs.
Here, we focus on the comparison of computational performance, which is shown in Fig.~\ref{Fig:performance}.
The test problem used here is a typical situation where structures with various sizes are subject to dynamic uniaxial tensile loadings.
We run DNS for continuum CMMs of sizes from 2$\times$2 to 8$\times$8 with~\texttt{FEniCS} on with a 2.4 GHz 8-Core Intel Core i9 CPU and record wall-clock time per integration step.
For DNS, a structure with size bigger than 8$\times$8 array of unit cells is already difficult to simulate.
For MGN based simulation, we consider sizes from 2$\times$2 up to 256$\times$256 array of unit cells.
We run the experiments on the same CPU mentioned above, and on an NVIDIA P100 GPU.
One great advantage of programming in in~\texttt{JAX} is that the code is designed for GPU, and when GPU is not available, the same code will run on CPU.
Each data point in Fig.~\ref{Fig:performance} is an average of 5 repeated experiments.
As observed in Fig.~\ref{Fig:performance}, the computational time required by MGN based approach is at least 2 orders of magnitude smaller than that by DNS.

The main reason of such great reduction in computational cost is the fact that MGN based approach uses a simplified cross-spring system where the DOFs involved in the numerical calculation are orders of magnitude fewer than DNS.
In addition, since the code can easily run on GPU with the help of~\texttt{JAX}, MGN with GPU support demonstrates even better computational efficiency.


\subsection{Shear wave propagation through large-size CMMs}
\label{subsec:HMM}

\begin{figure}[!ht]
    \centering 
    \scalebox{0.53}{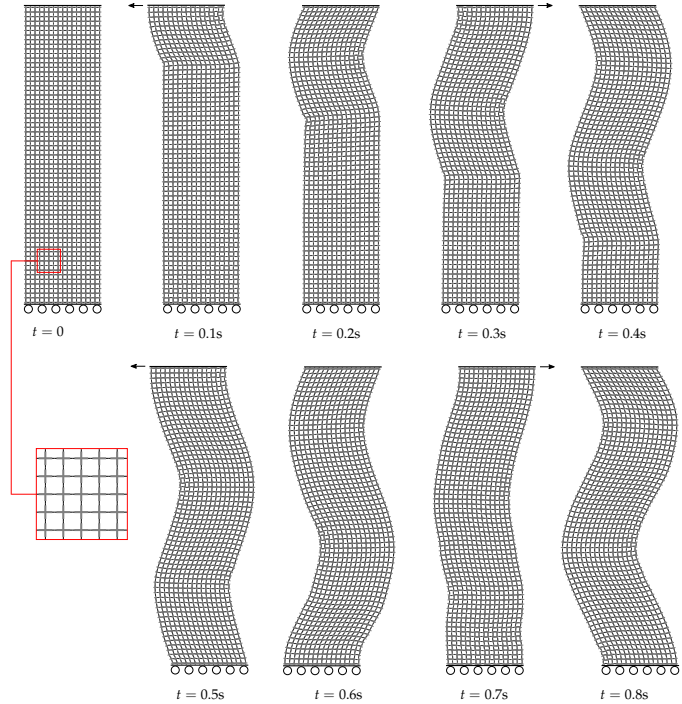}
    \caption{Shear wave propagation in a 16$\times$64 fully connected cross-spring system by MGN simulation for shape A.}
    \label{Fig:S_wave_hollow_0}
\end{figure}

In this final example, we consider CMMs whose sizes are relatively large and are practically impossible for DNS, but feasible with MGN.
We consider a 16$\times$64 cross-spring system, as shown in Fig.~\ref{Fig:S_wave_hollow_0}.
The structure is at rest initially and the following displacement is applied on the top side of the structure:
\begin{equation}
\label{Eq:shear-BC}
U_1(X_1, 64L_0, t) = 3.2L_0\textrm{sin}\left(-\frac{2\pi}{T}t\right), 
\end{equation}
where $T=0.4$s.
The amplitude $3.2L_0$ corresponds to $5\%$ of the total height of the structure.
Roller boundary condition is prescribed at the bottom to prevent vertical movement.
All other parts of the boundary are traction-free.

\begin{figure}[!ht]
    \centering 
    \scalebox{0.53}{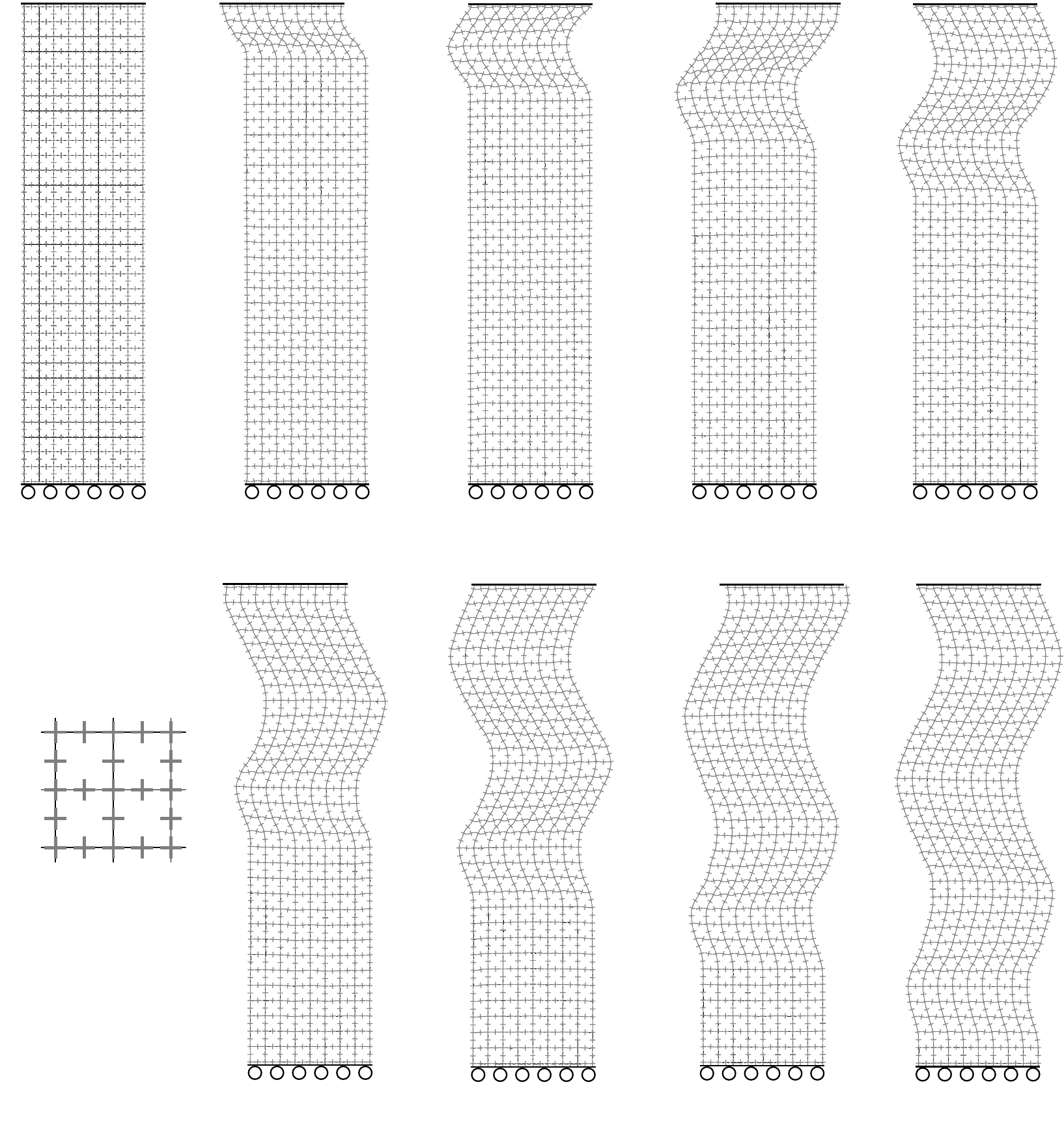}
    \caption{Shear wave propagation in a 16$\times$64 partially connected cross-spring system by MGN simulation for shape A.}
    \label{Fig:S_wave_hollow_2}
\end{figure}

\begin{figure}[!ht]
    \centering 
    \includegraphics[width=0.45\textwidth]{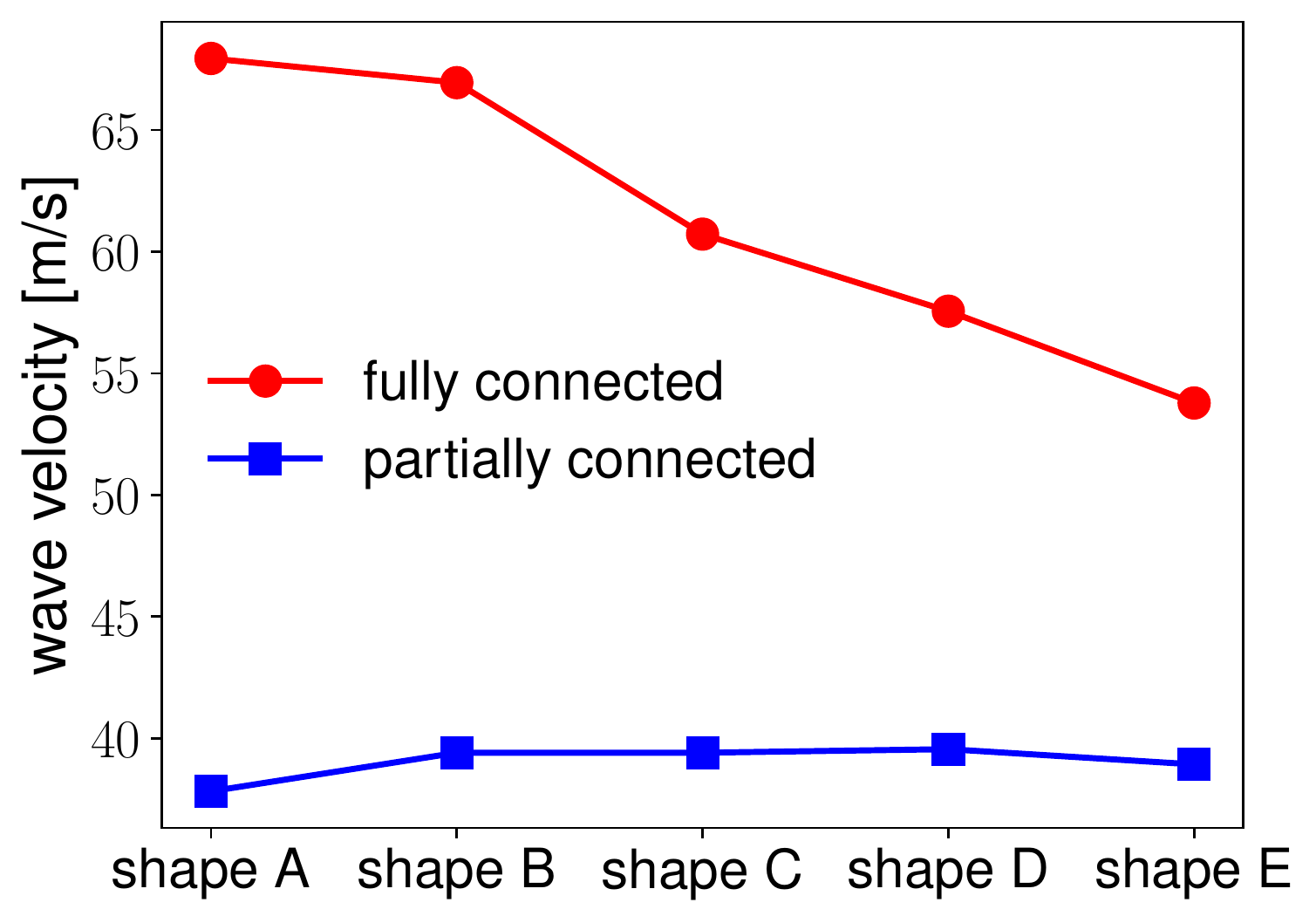}
    \caption{Shear wave speed for fully connected (red) and partially connected (blue) cross-spring systems with different pore shapes.}
    \label{Fig:S_wave}
\end{figure}

We perform this numerical experiment for all the five shapes, and only show results of shape A in Fig.~\ref{Fig:S_wave_hollow_0}.
From the figure we observe that the wave front propagates to the bottom side between $t=0.4$s and $t=0.5$s.
We roughly measure the shear wave velocity by tracking the onset of movement for the bottom row of crosses, and the corresponding velocities for the five pore shapes are shown in Fig.~\ref{Fig:S_wave} with the red color.
We observe a decreasing trend of wave velocities from shape A to shape E.
For reference, the same numerical experiment is performed on a continuum bulk material (zero porosity) with the same size, and the wave velocity is $213.3$m/s.


For perfect CMMs, it can be represented by a fully connected cross-spring system as shown in Fig.~\ref{Fig:S_wave_hollow_0} (see the red window).
By simply making the network a partially connected one, defects such as breakage can be easily incorporated in our MGN based approach. 
In Fig.~\ref{Fig:S_wave_hollow_2}, we introduce such defects in a periodic fashion, making it a hierarchical cross-spring system.
We apply the same loading conditions to such structure, which is of the same size as the previous example.
Eight snapshots of the deformed configurations are shown for the defected CMM with pore shape A.
The deformed configurations are quite different from the results of the perfect CMM with pore shape A and the deformation can be highly localized.
We studied the propagation of shear waves in defected CMMs with four other pore shapes as well and the measured wave velocities are reported in Fig.~\ref{Fig:S_wave}.
We can observe that the velocities of these defected structures are consistently lower than that of the corresponding perfect CMMs. 
In the meantime, the velocities are much less sensitive to pore shapes.


\section{Conclusion}
\label{Sec:conc}
In this work, we proposed a graph network based approach to predict the dynamic response of soft mechanical metamaterials. 
We start with a widely-used discrete cross-spring system as the simplified model of the elastodynamics of the continuum metamaterial.
We then use a metamaterial graph network to represent such system.
The nodal features of the network correspond to the positions and orientations of the rigid crosses and the edge update functions determine the elastic energy stored in the springs connecting neighbouring crosses.
We employ a data-driven approach to learn the elastic energy of the cross-spring system with different cellular geometries.
A database is built by carrying out finite element calculation over the corresponding building block of the metamaterial and is then used to train a model that best approximates the nonlinear mechanics of the metamaterial at the unit-cell scale.
Results show that Gaussian process regression is a better surrogate for constructing such model when compared to neural networks.
We then deploy the trained surrogate model to our metamaterial graph network to study the mechanics of the metamaterial at the structural scale with three representative examples.
In the first two examples, we compare results of the proposed approach to those produced by direct numerical simulation, where we observed a comparable accuracy, yet our proposed approach can significantly reduce the computational time.
In the final example, we use our proposed approach to study the dynamics of large size metamaterials that is practically not possible via direct numerical simulation.
We also demonstrate how defects and spatial inhomogeneities can be easily incorporated into our approach.
Through these examples, we demonstrate that the proposed approach is a fast yet reliable tool for the dynamic response of cellular mechanical materials with large sizes, which can have great potentials for developing efficient rational design algorithms of soft  mechanical metamaterials in the future.

\section*{Acknowledgements}
\label{Sec:Acknowledgements}

Sheng Mao acknowledges the financial support from the BIC-ESAT.




\bibliographystyle{plainnat}
\bibliography{refs}

\begin{thebibliography}{65}
\providecommand{\natexlab}[1]{#1}
\providecommand{\url}[1]{\texttt{#1}}
\expandafter\ifx\csname urlstyle\endcsname\relax
  \providecommand{\doi}[1]{doi: #1}\else
  \providecommand{\doi}{doi: \begingroup \urlstyle{rm}\Url}\fi

\bibitem[Barchiesi et~al.(2019)Barchiesi, Spagnuolo, and
  Placidi]{barchiesi2019mechanical}
Emilio Barchiesi, Mario Spagnuolo, and Luca Placidi.
\newblock Mechanical metamaterials: a state of the art.
\newblock \emph{Mathematics and Mechanics of Solids}, 24\penalty0 (1):\penalty0
  212--234, 2019.

\bibitem[Battaglia et~al.(2016)Battaglia, Pascanu, Lai, Rezende, and
  Kavukcuoglu]{battaglia2016interaction}
Peter~W Battaglia, Razvan Pascanu, Matthew Lai, Danilo Rezende, and Koray
  Kavukcuoglu.
\newblock Interaction networks for learning about objects, relations and
  physics.
\newblock \emph{arXiv preprint arXiv:1612.00222}, 2016.

\bibitem[Battaglia et~al.(2018)Battaglia, Hamrick, Bapst, Sanchez-Gonzalez,
  Zambaldi, Malinowski, Tacchetti, Raposo, Santoro, Faulkner,
  et~al.]{battaglia2018relational}
Peter~W Battaglia, Jessica~B Hamrick, Victor Bapst, Alvaro Sanchez-Gonzalez,
  Vinicius Zambaldi, Mateusz Malinowski, Andrea Tacchetti, David Raposo, Adam
  Santoro, Ryan Faulkner, et~al.
\newblock Relational inductive biases, deep learning, and graph networks.
\newblock \emph{arXiv preprint arXiv:1806.01261}, 2018.

\bibitem[Bertoldi(2017)]{bertoldi2017harnessing}
Katia Bertoldi.
\newblock Harnessing instabilities to design tunable architected cellular
  materials.
\newblock \emph{Annual Review of Materials Research}, 47:\penalty0 51--61,
  2017.

\bibitem[Bertoldi and Boyce(2008)]{bertoldi2008wave}
Katia Bertoldi and Mary~C Boyce.
\newblock Wave propagation and instabilities in monolithic and periodically
  structured elastomeric materials undergoing large deformations.
\newblock \emph{Physical Review B}, 78\penalty0 (18):\penalty0 184107, 2008.

\bibitem[Bertoldi et~al.(2010)Bertoldi, Reis, Willshaw, and
  Mullin]{bertoldi2010negative}
Katia Bertoldi, Pedro~M Reis, Stephen Willshaw, and Tom Mullin.
\newblock Negative poisson's ratio behavior induced by an elastic instability.
\newblock \emph{Advanced materials}, 22\penalty0 (3):\penalty0 361--366, 2010.

\bibitem[Bertoldi et~al.(2017)Bertoldi, Vitelli, Christensen, and van
  Hecke]{bertoldi2017flexible}
Katia Bertoldi, Vincenzo Vitelli, Johan Christensen, and Martin van Hecke.
\newblock Flexible mechanical metamaterials.
\newblock \emph{Nature Reviews Materials}, 2\penalty0 (11):\penalty0 17066,
  2017.

\bibitem[Bishop(2006)]{bishop2006pattern}
Christopher~M Bishop.
\newblock Pattern recognition.
\newblock \emph{Machine learning}, 128\penalty0 (9), 2006.

\bibitem[Bordes et~al.(2013)Bordes, Usunier, Garcia-Duran, Weston, and
  Yakhnenko]{bordes2013translating}
Antoine Bordes, Nicolas Usunier, Alberto Garcia-Duran, Jason Weston, and Oksana
  Yakhnenko.
\newblock Translating embeddings for modeling multi-relational data.
\newblock \emph{Advances in neural information processing systems}, 26, 2013.

\bibitem[Bradbury et~al.(2018)Bradbury, Frostig, Hawkins, Johnson, Leary,
  Maclaurin, Necula, Paszke, Vander{P}las, Wanderman-{M}ilne, and
  Zhang]{jax2018github}
James Bradbury, Roy Frostig, Peter Hawkins, Matthew~James Johnson, Chris Leary,
  Dougal Maclaurin, George Necula, Adam Paszke, Jake Vander{P}las, Skye
  Wanderman-{M}ilne, and Qiao Zhang.
\newblock {JAX}: composable transformations of {P}ython+{N}um{P}y programs,
  2018.
\newblock URL \url{http://github.com/google/jax}.

\bibitem[Byrd et~al.(1995)Byrd, Lu, Nocedal, and Zhu]{byrd1995limited}
Richard~H Byrd, Peihuang Lu, Jorge Nocedal, and Ciyou Zhu.
\newblock A limited memory algorithm for bound constrained optimization.
\newblock \emph{SIAM Journal on scientific computing}, 16\penalty0
  (5):\penalty0 1190--1208, 1995.

\bibitem[Chang et~al.(2016)Chang, Ullman, Torralba, and
  Tenenbaum]{chang2016compositional}
Michael~B Chang, Tomer Ullman, Antonio Torralba, and Joshua~B Tenenbaum.
\newblock A compositional object-based approach to learning physical dynamics.
\newblock \emph{arXiv preprint arXiv:1612.00341}, 2016.

\bibitem[Deng et~al.(2017)Deng, Raney, Tournat, and Bertoldi]{deng2017elastic}
Bolei Deng, JR~Raney, Vincent Tournat, and Katia Bertoldi.
\newblock Elastic vector solitons in soft architected materials.
\newblock \emph{Physical review letters}, 118\penalty0 (20):\penalty0 204102,
  2017.

\bibitem[Deng et~al.(2019{\natexlab{a}})Deng, Mo, Tournat, Bertoldi, and
  Raney]{deng2019focusing}
Bolei Deng, Chengyang Mo, Vincent Tournat, Katia Bertoldi, and Jordan~R Raney.
\newblock Focusing and mode separation of elastic vector solitons in a 2d soft
  mechanical metamaterial.
\newblock \emph{Physical review letters}, 123\penalty0 (2):\penalty0 024101,
  2019{\natexlab{a}}.

\bibitem[Deng et~al.(2019{\natexlab{b}})Deng, Zhang, He, Tournat, Wang, and
  Bertoldi]{deng2019propagation}
Bolei Deng, Yuning Zhang, Qi~He, Vincent Tournat, Pai Wang, and Katia Bertoldi.
\newblock Propagation of elastic solitons in chains of pre-deformed beams.
\newblock \emph{New Journal of Physics}, 21\penalty0 (7):\penalty0 073008,
  2019{\natexlab{b}}.

\bibitem[Deng et~al.(2021{\natexlab{a}})Deng, Li, Tournat, Purohit, and
  Bertoldi]{deng2021dynamics}
Bolei Deng, Jian Li, Vincent Tournat, Prashant~K Purohit, and Katia Bertoldi.
\newblock Dynamics of mechanical metamaterials: A framework to connect phonons,
  nonlinear periodic waves and solitons.
\newblock \emph{Journal of the Mechanics and Physics of Solids}, 147:\penalty0
  104233, 2021{\natexlab{a}}.

\bibitem[Deng et~al.(2021{\natexlab{b}})Deng, Raney, Bertoldi, and
  Tournat]{deng2021nonlinear}
Bolei Deng, JR~Raney, K~Bertoldi, and Vincent Tournat.
\newblock Nonlinear waves in flexible mechanical metamaterials.
\newblock \emph{Journal of Applied Physics}, 130\penalty0 (4):\penalty0 040901,
  2021{\natexlab{b}}.

\bibitem[Duvenaud et~al.(2015)Duvenaud, Maclaurin, Aguilera-Iparraguirre,
  G{\'o}mez-Bombarelli, Hirzel, Aspuru-Guzik, and
  Adams]{duvenaud2015convolutional}
David Duvenaud, Dougal Maclaurin, Jorge Aguilera-Iparraguirre, Rafael
  G{\'o}mez-Bombarelli, Timothy Hirzel, Al{\'a}n Aspuru-Guzik, and Ryan~P
  Adams.
\newblock Convolutional networks on graphs for learning molecular fingerprints.
\newblock \emph{arXiv preprint arXiv:1509.09292}, 2015.

\bibitem[Florijn et~al.(2016)Florijn, Coulais, and van
  Hecke]{florijn2016programmable}
Bastiaan Florijn, Corentin Coulais, and Martin van Hecke.
\newblock Programmable mechanical metamaterials: the role of geometry.
\newblock \emph{Soft Matter}, 12\penalty0 (42):\penalty0 8736--8743, 2016.

\bibitem[Goldsberry and Haberman(2018)]{goldsberry2018negative}
Benjamin~M Goldsberry and Michael~R Haberman.
\newblock Negative stiffness honeycombs as tunable elastic metamaterials.
\newblock \emph{Journal of Applied Physics}, 123\penalty0 (9):\penalty0 091711,
  2018.

\bibitem[Greydanus et~al.(2019)Greydanus, Dzumba, and
  Yosinski]{greydanus2019hamiltonian}
Samuel~J Greydanus, Misko Dzumba, and Jason Yosinski.
\newblock Hamiltonian neural networks.
\newblock 2019.

\bibitem[Hastie et~al.(2009)Hastie, Tibshirani, and
  Friedman]{hastie2009elements}
Trevor Hastie, Robert Tibshirani, and Jerome Friedman.
\newblock \emph{The elements of statistical learning: data mining, inference,
  and prediction}.
\newblock Springer Science \& Business Media, 2009.

\bibitem[Heider et~al.(2020)Heider, Wang, and Sun]{heider2020so}
Yousef Heider, Kun Wang, and WaiChing Sun.
\newblock So (3)-invariance of informed-graph-based deep neural network for
  anisotropic elastoplastic materials.
\newblock \emph{Computer Methods in Applied Mechanics and Engineering},
  363:\penalty0 112875, 2020.

\bibitem[Hoerl and Kennard(1970)]{hoerl1970ridge}
Arthur~E Hoerl and Robert~W Kennard.
\newblock Ridge regression: Biased estimation for nonorthogonal problems.
\newblock \emph{Technometrics}, 12\penalty0 (1):\penalty0 55--67, 1970.

\bibitem[Hughes(2012)]{hughes2012finite}
Thomas~JR Hughes.
\newblock \emph{The finite element method: linear static and dynamic finite
  element analysis}.
\newblock Courier Corporation, 2012.

\bibitem[Hussein et~al.(2014)Hussein, Leamy, and Ruzzene]{hussein2014dynamics}
Mahmoud~I Hussein, Michael~J Leamy, and Massimo Ruzzene.
\newblock Dynamics of phononic materials and structures: Historical origins,
  recent progress, and future outlook.
\newblock \emph{Applied Mechanics Reviews}, 66\penalty0 (4), 2014.

\bibitem[Kingma and Ba(2014)]{kingma2014adam}
Diederik~P Kingma and Jimmy Ba.
\newblock Adam: A method for stochastic optimization.
\newblock \emph{arxiv: 1412.6980}, 2014.

\bibitem[Kochkov et~al.(2021)Kochkov, Smith, Alieva, Wang, Brenner, and
  Hoyer]{kochkov2021machine}
Dmitrii Kochkov, Jamie~A Smith, Ayya Alieva, Qing Wang, Michael~P Brenner, and
  Stephan Hoyer.
\newblock Machine learning--accelerated computational fluid dynamics.
\newblock \emph{Proceedings of the National Academy of Sciences}, 118\penalty0
  (21), 2021.

\bibitem[Kochmann and Bertoldi(2017)]{kochmann2017exploiting}
Dennis~M Kochmann and Katia Bertoldi.
\newblock Exploiting microstructural instabilities in solids and structures:
  from metamaterials to structural transitions.
\newblock \emph{Applied mechanics reviews}, 69\penalty0 (5), 2017.

\bibitem[Krishnan and Johnson(2009)]{krishnan2009optical}
D~Krishnan and Harley~T Johnson.
\newblock Optical properties of two-dimensional polymer photonic crystals after
  deformation-induced pattern transformations.
\newblock \emph{Journal of the Mechanics and Physics of Solids}, 57\penalty0
  (9):\penalty0 1500--1513, 2009.

\bibitem[Li and Gao(2016)]{li2016mechanical}
Xiaoyan Li and Huajian Gao.
\newblock Mechanical metamaterials: Smaller and stronger.
\newblock \emph{Nature Materials}, 15\penalty0 (4):\penalty0 373, 2016.

\bibitem[Li et~al.(2017)Li, Yu, Shahabi, and Liu]{li2017diffusion}
Yaguang Li, Rose Yu, Cyrus Shahabi, and Yan Liu.
\newblock Diffusion convolutional recurrent neural network: Data-driven traffic
  forecasting.
\newblock \emph{arXiv preprint arXiv:1707.01926}, 2017.

\bibitem[Logg et~al.(2012)Logg, Mardal, and Wells]{logg2012automated}
Anders Logg, Kent-Andre Mardal, and Garth Wells.
\newblock \emph{Automated solution of differential equations by the finite
  element method: The FEniCS book}, volume~84.
\newblock Springer Science \& Business Media, 2012.

\bibitem[Marsden and Hughes(1994)]{marsden1994mathematical}
Jerrold~E Marsden and Thomas~JR Hughes.
\newblock \emph{Mathematical foundations of elasticity}.
\newblock Courier Corporation, 1994.

\bibitem[Meza et~al.(2014)Meza, Das, and Greer]{meza2014strong}
Lucas~R Meza, Satyajit Das, and Julia~R Greer.
\newblock Strong, lightweight, and recoverable three-dimensional ceramic
  nanolattices.
\newblock \emph{Science}, 345\penalty0 (6202):\penalty0 1322--1326, 2014.

\bibitem[Mirzaali et~al.(2018)Mirzaali, Janbaz, Strano, Vergani, and
  Zadpoor]{mirzaali2018shape}
MJ~Mirzaali, Shahram Janbaz, M~Strano, L~Vergani, and Amir~A Zadpoor.
\newblock Shape-matching soft mechanical metamaterials.
\newblock \emph{Scientific reports}, 8\penalty0 (1):\penalty0 965, 2018.

\bibitem[Mitchell(1980)]{mitchell1980need}
Tom~M Mitchell.
\newblock \emph{The need for biases in learning generalizations}.
\newblock Department of Computer Science, Laboratory for Computer Science
  Research~…, 1980.

\bibitem[Monaghan(1992)]{monaghan1992smoothed}
Joe~J Monaghan.
\newblock Smoothed particle hydrodynamics.
\newblock \emph{Annual review of astronomy and astrophysics}, 30\penalty0
  (1):\penalty0 543--574, 1992.

\bibitem[Mozaffar et~al.(2021)Mozaffar, Liao, Lin, Ehmann, and
  Cao]{mozaffar2021geometry}
Mojtaba Mozaffar, Shuheng Liao, Hui Lin, Kornel Ehmann, and Jian Cao.
\newblock Geometry-agnostic data-driven thermal modeling of additive
  manufacturing processes using graph neural networks.
\newblock \emph{Additive Manufacturing}, 48:\penalty0 102449, 2021.

\bibitem[Nadkarni et~al.(2014)Nadkarni, Daraio, and
  Kochmann]{nadkarni2014dynamics}
Neel Nadkarni, Chiara Daraio, and Dennis~M Kochmann.
\newblock Dynamics of periodic mechanical structures containing bistable
  elastic elements: From elastic to solitary wave propagation.
\newblock \emph{Physical Review E}, 90\penalty0 (2):\penalty0 023204, 2014.

\bibitem[Ne{\v{c}}as(1974)]{nevcas1974application}
Jind{\v{r}}ich Ne{\v{c}}as.
\newblock Application of rothe's method to abstract parabolic equations.
\newblock \emph{Czechoslovak Mathematical Journal}, 24\penalty0 (3):\penalty0
  496--500, 1974.

\bibitem[Nesterenko(2013)]{nesterenko2013dynamics}
Vitali Nesterenko.
\newblock \emph{Dynamics of heterogeneous materials}.
\newblock Springer Science \& Business Media, 2013.

\bibitem[Ogden(1997)]{ogden1997non}
Raymond~W Ogden.
\newblock \emph{Non-linear elastic deformations}.
\newblock Courier Corporation, 1997.

\bibitem[Overvelde and Bertoldi(2014)]{overvelde2014relating}
Johannes~TB Overvelde and Katia Bertoldi.
\newblock Relating pore shape to the non-linear response of periodic
  elastomeric structures.
\newblock \emph{Journal of the Mechanics and Physics of Solids}, 64:\penalty0
  351--366, 2014.

\bibitem[Overvelde et~al.(2012)Overvelde, Shan, and
  Bertoldi]{overvelde2012compaction}
Johannes~TB Overvelde, Sicong Shan, and Katia Bertoldi.
\newblock Compaction through buckling in 2d periodic, soft and porous
  structures: effect of pore shape.
\newblock \emph{Advanced Materials}, 24\penalty0 (17):\penalty0 2337--2342,
  2012.

\bibitem[Pence and Gou(2015)]{pence2015compressible}
Thomas~J Pence and Kun Gou.
\newblock On compressible versions of the incompressible neo-hookean material.
\newblock \emph{Mathematics and Mechanics of Solids}, 20\penalty0 (2):\penalty0
  157--182, 2015.

\bibitem[Pfaff et~al.(2020)Pfaff, Fortunato, Sanchez-Gonzalez, and
  Battaglia]{pfaff2020learning}
Tobias Pfaff, Meire Fortunato, Alvaro Sanchez-Gonzalez, and Peter~W Battaglia.
\newblock Learning mesh-based simulation with graph networks.
\newblock \emph{arXiv preprint arXiv:2010.03409}, 2020.

\bibitem[Press et~al.(1986)Press, Vetterling, Teukolsky, and
  Flannery]{press1986numerical}
William~H Press, William~T Vetterling, Saul~A Teukolsky, and Brian~P Flannery.
\newblock \emph{Numerical recipes}, volume 818.
\newblock Cambridge university press Cambridge, 1986.

\bibitem[Rasmussen(2003)]{rasmussen2003gaussian}
Carl~Edward Rasmussen.
\newblock Gaussian processes in machine learning.
\newblock In \emph{Summer school on machine learning}, pages 63--71. Springer,
  2003.

\bibitem[Rektorys(1971)]{rektorys1971application}
Karel Rektorys.
\newblock On application of direct variational methods to the solution of
  parabolic boundary value problems of arbitrary order in the space variables.
\newblock \emph{Czechoslovak Mathematical Journal}, 21\penalty0 (2):\penalty0
  318--339, 1971.

\bibitem[Rothe(1930)]{rothe1930zweidimensionale}
Erich Rothe.
\newblock Zweidimensionale parabolische randwertaufgaben als grenzfall
  eindimensionaler randwertaufgaben.
\newblock \emph{Mathematische Annalen}, 102\penalty0 (1):\penalty0 650--670,
  1930.

\bibitem[Sanchez-Gonzalez et~al.(2018)Sanchez-Gonzalez, Heess, Springenberg,
  Merel, Riedmiller, Hadsell, and Battaglia]{sanchez2018graph}
Alvaro Sanchez-Gonzalez, Nicolas Heess, Jost~Tobias Springenberg, Josh Merel,
  Martin Riedmiller, Raia Hadsell, and Peter Battaglia.
\newblock Graph networks as learnable physics engines for inference and
  control.
\newblock In \emph{International Conference on Machine Learning}, pages
  4470--4479. PMLR, 2018.

\bibitem[Sanchez-Gonzalez et~al.(2020)Sanchez-Gonzalez, Godwin, Pfaff, Ying,
  Leskovec, and Battaglia]{sanchez2020learning}
Alvaro Sanchez-Gonzalez, Jonathan Godwin, Tobias Pfaff, Rex Ying, Jure
  Leskovec, and Peter Battaglia.
\newblock Learning to simulate complex physics with graph networks.
\newblock In \emph{International Conference on Machine Learning}, pages
  8459--8468. PMLR, 2020.

\bibitem[Santoro et~al.(2017)Santoro, Raposo, Barrett, Malinowski, Pascanu,
  Battaglia, and Lillicrap]{santoro2017simple}
Adam Santoro, David Raposo, David~GT Barrett, Mateusz Malinowski, Razvan
  Pascanu, Peter Battaglia, and Timothy Lillicrap.
\newblock A simple neural network module for relational reasoning.
\newblock \emph{arXiv preprint arXiv:1706.01427}, 2017.

\bibitem[Schoenholz and Cubuk(2020)]{schoenholz2020jax}
Samuel Schoenholz and Ekin~Dogus Cubuk.
\newblock Jax md: a framework for differentiable physics.
\newblock \emph{Advances in Neural Information Processing Systems}, 33, 2020.

\bibitem[Shan et~al.(2014)Shan, Kang, Wang, Qu, Shian, Chen, and
  Bertoldi]{shan2014harnessing}
Sicong Shan, Sung~H Kang, Pai Wang, Cangyu Qu, Samuel Shian, Elizabeth~R Chen,
  and Katia Bertoldi.
\newblock Harnessing multiple folding mechanisms in soft periodic structures
  for tunable control of elastic waves.
\newblock \emph{Advanced Functional Materials}, 24\penalty0 (31):\penalty0
  4935--4942, 2014.

\bibitem[Sulsky et~al.(1995)Sulsky, Zhou, and Schreyer]{sulsky1995application}
Deborah Sulsky, Shi-Jian Zhou, and Howard~L Schreyer.
\newblock Application of a particle-in-cell method to solid mechanics.
\newblock \emph{Computer physics communications}, 87\penalty0 (1-2):\penalty0
  236--252, 1995.

\bibitem[Surjadi et~al.(2019)Surjadi, Gao, Du, Li, Xiong, Fang, and
  Lu]{surjadi2019mechanical}
James~Utama Surjadi, Libo Gao, Huifeng Du, Xiang Li, Xiang Xiong,
  Nicholas~Xuanlai Fang, and Yang Lu.
\newblock Mechanical metamaterials and their engineering applications.
\newblock \emph{Advanced Engineering Materials}, 21\penalty0 (3):\penalty0
  1800864, 2019.

\bibitem[Theocharis et~al.(2013)Theocharis, Boechler, and
  Daraio]{theocharis2013nonlinear}
G~Theocharis, N~Boechler, and C~Daraio.
\newblock Nonlinear periodic phononic structures and granular crystals.
\newblock In \emph{Acoustic Metamaterials and Phononic Crystals}, pages
  217--251. Springer, 2013.

\bibitem[Tibshirani(1996)]{tibshirani1996regression}
Robert Tibshirani.
\newblock Regression shrinkage and selection via the lasso.
\newblock \emph{Journal of the Royal Statistical Society: Series B
  (Methodological)}, 58\penalty0 (1):\penalty0 267--288, 1996.

\bibitem[Tournat and Gusev(2010)]{tournat2010acoustics}
V~Tournat and VE~Gusev.
\newblock Acoustics of unconsolidated “model” granular media: An overview
  of recent results and several open problems.
\newblock \emph{Acta Acustica united with Acustica}, 96\penalty0 (2):\penalty0
  208--224, 2010.

\bibitem[Vlassis et~al.(2020)Vlassis, Ma, and Sun]{vlassis2020geometric}
Nikolaos~N Vlassis, Ran Ma, and WaiChing Sun.
\newblock Geometric deep learning for computational mechanics part i:
  Anisotropic hyperelasticity.
\newblock \emph{Computer Methods in Applied Mechanics and Engineering},
  371:\penalty0 113299, 2020.

\bibitem[Xue et~al.(2020)Xue, Beatson, Chiaramonte, Roeder, Ash, Menguc,
  Adriaenssens, Adams, and Mao]{xue2020data}
Tianju Xue, Alex Beatson, Maurizio Chiaramonte, Geoffrey Roeder, Jordan~T Ash,
  Yigit Menguc, Sigrid Adriaenssens, Ryan~P Adams, and Sheng Mao.
\newblock A data-driven computational scheme for the nonlinear mechanical
  properties of cellular mechanical metamaterials under large deformation.
\newblock \emph{Soft matter}, 16\penalty0 (32):\penalty0 7524--7534, 2020.

\bibitem[Yasuda et~al.(2019)Yasuda, Miyazawa, Charalampidis, Chong, Kevrekidis,
  and Yang]{yasuda2019origami}
Hiromi Yasuda, Yasuhiro Miyazawa, Efstathios~G Charalampidis, Christopher
  Chong, Panayotis~G Kevrekidis, and Jinkyu Yang.
\newblock Origami-based impact mitigation via rarefaction solitary wave
  creation.
\newblock \emph{Science advances}, 5\penalty0 (5):\penalty0 eaau2835, 2019.

\bibitem[Zhu et~al.(1997)Zhu, Byrd, Lu, and Nocedal]{zhu1997algorithm}
Ciyou Zhu, Richard~H Byrd, Peihuang Lu, and Jorge Nocedal.
\newblock Algorithm 778: L-bfgs-b: Fortran subroutines for large-scale
  bound-constrained optimization.
\newblock \emph{ACM Transactions on mathematical software (TOMS)}, 23\penalty0
  (4):\penalty0 550--560, 1997.

\end{thebibliography}

\appendix
\setcounter{figure}{0}

\section{Numerical discretization for the continuum problem}
\label{App:dns}

We introduce velocity $\Vb=\frac{\partial \Ub}{\partial t}$ as a separate variable and then Eq.~(\ref{Eq:BVP-strong}) can be re-written as

\begin{align} \label{Eq:uv-strong}
   \rho_R\frac{\partial \Vb}{\partial t}   = \textrm{Div }  \Pb &\quad\quad \textrm{in}  \, \,  \mathbb{B}\times [0, T] , \nonumber \\
    \frac{\partial \Ub}{\partial t} = \Vb  &\quad\quad\textrm{in} \, \, \mathbb{B} \times [0, T],  \nonumber \\
    \Ub(\X, 0)=\Ub_0  &\quad\quad\textrm{in} \, \, \mathbb{B},  \nonumber \\
    \Vb(\X, 0) =\Vb_0  &\quad\quad\textrm{in} \, \, \mathbb{B},  \nonumber \\
    \Ub = \Ub_D   &\quad\quad\textrm{on} \, \, \partial\mathbb{B}_D \times [0, T],  \nonumber \\
    \Pb \cdot \Nb = \Tb_N  &\quad\quad \textrm{on} \, \, \partial\mathbb{B}_N \times [0, T].
\end{align}

The above PDEs contain only first order time derivatives.
Let superscript $n$ denote the $n$th time step, we adopt the following scheme for temporal discretization:
\begin{align} \label{Eq:crank-nicolson}
    \rho_R\frac{\Vb^{n+1} - \Vb^{n}}{\Delta t} &= \textrm{Div } \Pb \big[ \alpha \Ub^{n+1} + (1-\alpha)\Ub^{n} \big],     \nonumber \\
    \frac{\Ub^{n+1} - \Ub^n}{\Delta t} &= \alpha \Vb^{n+1} + (1-\alpha)\Vb^n,
\end{align}
where $\Delta t$ is the time increment and we solve for $\Ub^{n+1}$ and $\Vb^{n+1}$ given $\Ub^{n}$ and $\Vb^{n}$, which are obtained from the previous time step. 
A coefficient $\alpha$ is introduced here to represent different schemes.
When $\alpha=0$, the scheme is explicit Euler; when $\alpha=1$, the scheme is implicit Euler.
In this work we set $\alpha=0.5$, which corresponds to the well-known Crank-Nicolson scheme.

Standard finite element method can be used to solve Eq.~(\ref{Eq:crank-nicolson}) at each time step.
We omit the details of deriving the weak form of Eq.~(\ref{Eq:crank-nicolson}) and  constructing the Newton-Raphson solver, since those procedures are automated by \texttt{FEniCS}.
Interested readers may refer to our code for full details of implementation. 

\section{Numerical integration for the cross-spring system}
\label{App:cs}

By introducing velocity $\vb_i= \frac{\textrm{d}\x_i}{\textrm{d}t}$, and angular velocity $\omega_i =  \frac{\textrm{d}\theta_i}{\textrm{d}t}$, Eq.~(\ref{Eq:strong_cs}) can be transformed to a system of first order ODEs:
\begin{align} \label{Eq:first_order_strong_cs}
    \frac{\textrm{d} \vb_i}{\textrm{d}t} &= -\frac{1}{m_i}\frac{\partial \Psi}{\partial \x_i} + \frac{\fb_i}{m_i},    \nonumber \\
    \frac{\textrm{d}\x_i}{\textrm{d}t} &= \vb_i,  \nonumber \\
    \frac{\textrm{d} \omega_i}{\textrm{d}t} &= -\frac{1}{I_i}\frac{\partial \Psi}{\partial \theta_i} + \frac{\tau_i}{I_i},  \nonumber \\
   \frac{\textrm{d}\theta_i}{\textrm{d}t} &= \omega_i,
\end{align}
where $i=1,2,...,N_c$ with some initial conditions prescribed for $\x_i$, $\vb_i$, $\theta_i$, and $\omega_i$.

Let superscript $n$ denote the $n$th time step.
We follow the well-known leap-frog scheme~\citep{press1986numerical} to carry out the numerical integration.
In such scheme, $\vb_i^{n-1/2}$, $\x_i^n$, $\omega_i^{n-1/2}$, and $\theta_i^n$ are known from the previous step.
We first solve for $\vb_i^{n+1/2}$ and $\omega_i^{n+1/2}$, and then solve for $\x_i^{n+1}$ and $\theta_i^{n+1}$ based on the solutions:
\begin{align} \label{Eq:leapfrog}
    \frac{\vb_i^{n+1/2} - \vb_i^{n-1/2}}{\Delta t} &= -\frac{1}{m}\frac{\partial \Psi(\Xi^n, \Theta^n)}{\partial \x_i^n} + \frac{\fb_i^n}{m},  \nonumber \\
    \frac{\x_i^{n+1} - \x_i^{n}}{\Delta t} & = \vb_i^{n+1/2},  \nonumber \\
    \frac{\omega_i^{n+1/2} - \omega_i^{n-1/2}}{\Delta t} &= -\frac{1}{I}\frac{\partial \Psi(\Xi^n, \Theta^n)}{\partial \theta_i^n} + \frac{\tau_i^n}{I},  \nonumber \\
    \frac{\theta_i^{n+1} - \theta_i^{n}}{\Delta t} & = \omega_i^{n+1/2}. \nonumber \\
\end{align} 

The numerical integration is implemented is in \texttt{JAX}.

\section{Data preparation}
\label{App:data_prep}

The ``ground truth'' data is obtained via solving finite element simulation over the building block of the CMM.
Each data point contains a feature vector $(\tilde{\theta}_a, \tilde{\theta}_b, d)$ and an output energy scalar $\tilde{\psi}$.

\begin{figure}[H]
    \centering 
    \scalebox{1}{
\begingroup%
  \makeatletter%
  \providecommand\color[2][]{%
    \errmessage{(Inkscape) Color is used for the text in Inkscape, but the package 'color.sty' is not loaded}%
    \renewcommand\color[2][]{}%
  }%
  \providecommand\transparent[1]{%
    \errmessage{(Inkscape) Transparency is used (non-zero) for the text in Inkscape, but the package 'transparent.sty' is not loaded}%
    \renewcommand\transparent[1]{}%
  }%
  \providecommand\rotatebox[2]{#2}%
  \newcommand*\fsize{\dimexpr\f@size pt\relax}%
  \newcommand*\lineheight[1]{\fontsize{\fsize}{#1\fsize}\selectfont}%
  \ifx\svgwidth\undefined%
    \setlength{\unitlength}{268.0247711bp}%
    \ifx\svgscale\undefined%
      \relax%
    \else%
      \setlength{\unitlength}{\unitlength * \real{\svgscale}}%
    \fi%
  \else%
    \setlength{\unitlength}{\svgwidth}%
  \fi%
  \global\let\svgwidth\undefined%
  \global\let\svgscale\undefined%
  \makeatother%
  \begin{picture}(1,0.70297821)%
    \lineheight{1}%
    \setlength\tabcolsep{0pt}%
    \put(0,0){\includegraphics[width=\unitlength,page=1]{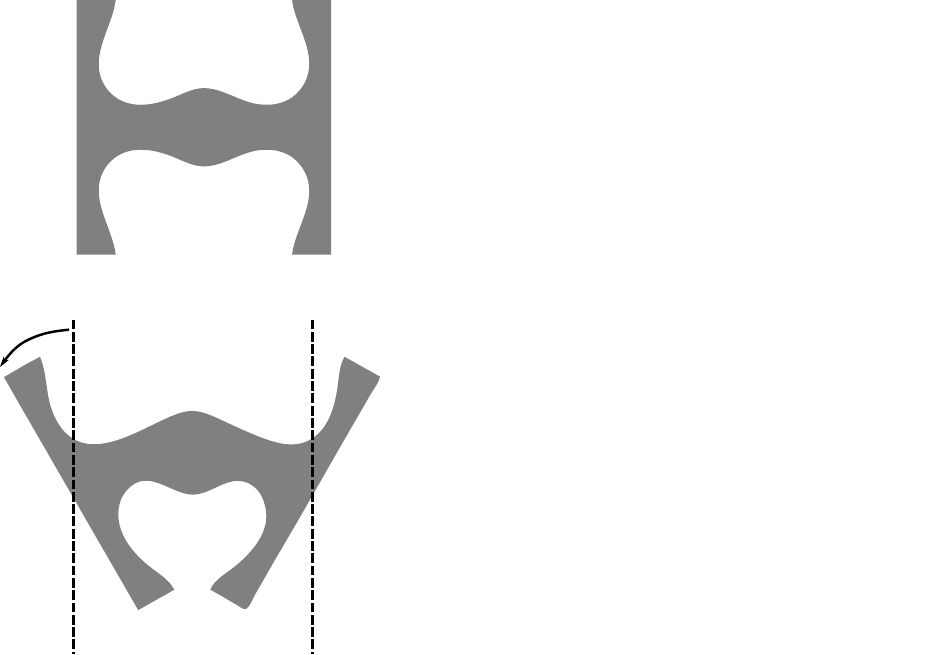}}%
    \put(0.00831633,0.35349805){\color[rgb]{0,0,0}\makebox(0,0)[lt]{\lineheight{1.25}\smash{\begin{tabular}[t]{l}$\tilde{\theta}_a$\end{tabular}}}}%
    \put(0,0){\includegraphics[width=\unitlength,page=2]{data_prep.pdf}}%
    \put(0.38348561,0.35205384){\color[rgb]{0,0,0}\makebox(0,0)[lt]{\lineheight{1.25}\smash{\begin{tabular}[t]{l}$\tilde{\theta}_b$\end{tabular}}}}%
    \put(0,0){\includegraphics[width=\unitlength,page=3]{data_prep.pdf}}%
    \put(0.28515196,0.39011746){\color[rgb]{0,0,0}\makebox(0,0)[lt]{\lineheight{1.25}\smash{\begin{tabular}[t]{l}$d$\end{tabular}}}}%
    \put(0,0){\includegraphics[width=\unitlength,page=4]{data_prep.pdf}}%
    \put(0.9746948,0.08463586){\color[rgb]{0,0,0}\makebox(0,0)[lt]{\lineheight{1.25}\smash{\begin{tabular}[t]{l}$W$\end{tabular}}}}%
    \put(0,0){\includegraphics[width=\unitlength,page=5]{data_prep.pdf}}%
  \end{picture}%
\endgroup%
}
    \caption{Settings of the local statics problem. The upper left part shows the continuum building block in its reference configuration. The lower left part shows the deformed building block in its spatial configuration. The lower right part shows the finite element mesh as well as the spatial distribution of the strain energy density $W$.}
    \label{Fig:data_prep}
\end{figure}

Let us take shape E as an example. 
For this process, we take a building block of the continuum CMM (shown in Fig.~\ref{Fig:data_prep}), and solve the following elastostatic problem 
\begin{align} \label{Eq:BVP-local}
    \textrm{Div }  \Pb = \textbf{0}  & \quad \textrm{in}  \, \, \mathbb{B} , \nonumber \\
    \Ub = \Ub_D &  \quad\textrm{on}  \, \, \partial\mathbb{B}_D,  \nonumber \\
    \Pb \cdot \Nb = \textbf{0} & \quad \textrm{on} \, \, \partial\mathbb{B}_N,
\end{align}
where $\mathbb{B}$ is the material domain of the building block in its reference configuration, $\mathbb{B}_D$ is labelled with red color, and $\mathbb{B}_N$ is the rest of the boundary.
As shown Fig.~\ref{Fig:data_prep}, the Dirichlet boundary conditions are imposed according the feature vector $(\tilde{\theta}_a, \tilde{\theta}_b, d)$. 
Namely, we rotate the left and right sides of the building block by specified angles $\tilde{\theta}_a$ and $\tilde{\theta}_b$, and also impose a stretch/compression according to the specified displacement $d$.
In this example, we have $\tilde{\theta}_a=-\frac{\pi}{6}$, $\tilde{\theta}_b=\frac{\pi}{6}$, and $d=-0.1L_0$.
Traction free conditions are imposed on $\partial \mathbb{B}_N$.
After solving this local boundary value problem, we compute the total internal potential energy as 
\begin{align}
    \tilde{\psi} = \int_{\mathbb{B}} W \textrm{ d}\X.
\end{align}
The strain energy density $W$ is shown in Fig.~\ref{Fig:data_prep}, where large values of $W$ are observed mostly in the middle ``ligament''.  
This observation helps to justify our assumption to put the total elastic energy of the building block into the connecting spring.

We draw 1000 samples from uniform distribution of a cube $(-\frac{\pi}{5}, \frac{\pi}{5}) \times (-\frac{\pi}{5}, \frac{\pi}{5}) \times (-0.2L_0, 0.2L_0)$ and solve the local problem for each sample to obtain a data set $D=\lbrace (\zb;y)^{(i)}\rbrace_{i=1:|D|}$, where $\zb:=(\tilde{\theta}_a, \tilde{\theta}_b, d)$ is the feature vector, $y:=\tilde{\psi}$ is the output energy scalar, and $|D|=1000$ is the size. 
This process is repeated for four other pore shapes (see Fig.~\ref{Fig:pore}).

\end{document}